\pdfoutput=1

\PassOptionsToPackage{table,xcdraw}{xcolor}
\PassOptionsToPackage{hyphens}{url}

\documentclass[a4paper,fleqn]{cas-sc}

\usepackage[authoryear,longnamesfirst]{natbib}

\usepackage{bm}
\usepackage{multicol}
\usepackage[algosection,linesnumbered,lined,boxed,commentsnumbered]{algorithm2e}
\usepackage{float}
\usepackage[caption=false]{subfig}

\input{supplement_labels}

\newcommand{\fpr}[1]{\left(#1\right)}
\newcommand{\tpr}[1]{\left[#1\right]}
\newcommand{\spr}[1]{\left\{#1\right\}}

\newcommand{\V}[1]{{\bm{\mathbf{\MakeLowercase{#1}}}}} 
\newcommand{\M}[1]{{\bm{\mathbf{\MakeUppercase{#1}}}}} 

\def\heta{h_{\eta}}

\def\bG{\mathbf{G}}
\def\bI{\mathbf{I}}
\def\bW{\mathbf{W}}

\def\bX{\mathbf{X}}
\def\bY{\mathbf{Y}}

\def\bZ{\mathbf{Z}}
\def\bP{\mathbf{P}}

\def\bV{\mathbf{V}}

\def\btW{\widetilde{\bW}}
\def\btH{\widetilde{\bH}}

\def\btX{\widetilde{\bX}}

\def\btZ{\widetilde{\bZ}}
\def\btP{\widetilde{\bP}}

\def\btV{\widetilde{\bV}}

\def\mcS{\mathcal{S}}

\def\mcD{\mathcal{D}}

\def\bY{\mathbf{Y}}
\def\bD{\mathbf{D}}

\def\sprime{s^{\prime}}

\def\bV{\mathbf{V}}

\def\heta{\widehat{\eta}}
\def\hgamma{\widehat{\gamma}}

\def\balpha{\boldsymbol{\alpha}}

\def\bSigma{\boldsymbol{\Sigma}}

\def\bhSigma{\widehat{\bSigma}}

\def\btH{\widetilde{\mathbf{H}}}

\DeclareMathOperator*{\argmin}{arg\,min}

\newtheorem{theorem}{Theorem}

\newtheorem{corollary}[theorem]{Corollary}

\begin{document}
\let\WriteBookmarks\relax
\def\floatpagepagefraction{1}
\def\textpagefraction{.001}

\shorttitle{Projection-based Test for Significance of Smooth Effect}
\shortauthors{Koner, Staicu, and Maity}

\title [mode = title]{Projection-based significance test for longitudinally collected functional data under a crossover design}

\author[1]{Salil Koner}[orcid=0000-0003-1952-4210]
\cormark[1]
\ead{salil.koner@ucr.edu}

\affiliation[1]{organization={Department of Statistics, University of California Riverside},
                city={Riverside},
                state={CA},
                country={USA}}

\author[2]{Ana-Maria Staicu}
\ead{astaicu@ncsu.edu}

\author[2]{Arnab Maity}
\ead{amaity@ncsu.edu}

\affiliation[2]{organization={Department of Statistics, North Carolina State University},
                city={Raleigh},
                state={NC},
                country={USA}}

\cortext[cor1]{Corresponding author. Salil Koner is an Assistant Professor of Statistics at University of California Riverside.}

\begin{abstract}
Wearable devices for continuous electronic health monitoring often capture data at frequent intervals under a dense functional design. The focal point is the analysis of longitudinal functional data, wherein functional trajectories are observed repeatedly over time. This work is motivated by the interest in assessing the efficacy of a noninflammatory medication, meloxicam, on the daily activity levels of household cats with a pre-existing condition of osteoarthritis under a crossover design. An accelerometer records these activity profiles at a minute level over the entire study period. To this aspect, we propose an orthogonal projection-based pseudo-generalized F test to determine the significance of the functional treatment effect under a functional additive crossover model while accounting for carryover effects and baseline covariates. Under mild conditions, we derive the asymptotic null distribution of the test statistic and the theoretical power function when the projection function is estimated from the data. Numerical studies demonstrate that the proposed test maintains size, proves powerful in detecting the smooth effect of meloxicam, and exhibits high efficiency compared to bootstrap-based alternatives. Application of the test on the accelerometric activity profiles reveals a strong positive effect of meloxicam on the joint pain score of the cats after adjusting for other baseline covariates.
\end{abstract}


\maketitle

\section{Introduction}

The prospect of monitoring health electronically has led to rapid use of wearable devices that are capable of collecting abundant amounts of health-related information continuously over time. Accelerometer is one of the most popular wearable devices that can objectively measure the physical activity (PA) count as densely as at a minute level \citep{bussmann2001measuring}. Most literature on this topic, summarizes the massive activity data using various summary measures in pursuit of explaining the association between activities and health outcomes \citep{reider2020methods}. However, such summaries-based approaches eliminate the variation of the PA over time. Functional data analysis (FDA)-based approaches view the daily PA counts as realization of some latent stochastic process and focus on modeling, prediction, and studying the association between response with various important covariates such as age, gender, etc. through functional mixed models. See \cite{zhang2019review} for a comprehensive review of the existing statistical methodologies on the accelerometric PA profiles. 

 Modeling daily PA involves handling complex functional dependence that inherently exists in the data. When the daily PA activities are recorded over multiple days, \cite{goldsmith2015generalized} proposed a multilevel functional data method to analyze PA profiles, which was extended by \cite{xiao2015quantifying} to account for subject-specific covariates. However, in many applications response trajectories are not observed daily for each subject, and the days over which they are observed are different for each subject.
In our motivating study, household cats were subjected to a placebo-controlled four-period crossover design \citep{ratkowsky1992cross}  with an intermediate washout period where daily PA profiles were measured at every minute level for each subject through accelerometer over 12 weeks (Table~\ref{tab: chap3crossoverlayout}). Such a data structure where functional response profiles are observed with a small number of curves per subject, falls under the framework of longitudinal functional data (LFD) \citep{di2009multilevel}. 
\begin{table}[pos=ht]
\centering
\caption{Layout of the four-period crossover design with active treatment meloxicam}
\label{tab: chap3crossoverlayout}
\begin{tabular}{|c|c|c|c|c|}
\hline
Group & \begin{tabular}[c]{@{}l@{}}Period 1\\ (3 weeks)\end{tabular} & \multicolumn{1}{l|}{\begin{tabular}[c]{@{}l@{}}Period 2\\ (3 weeks)\end{tabular}} & \begin{tabular}[c]{@{}l@{}}Period 3\\ (3 weeks)\end{tabular} & \begin{tabular}[c]{@{}l@{}}Period 4\\ (3 weeks)\end{tabular} \\ \hline
\multirow{2}{*}{1 (29 subjects)} & \multirow{2}{*}{\textcolor{red}{Meloxicam}} & \multirow{4}{*}{\textcolor{brown}{Washout}} & \multirow{2}{*}{\textcolor{blue}{Placebo}} & \multirow{4}{*}{\textcolor{brown}{Washout}} \\
 &  &  &  &  \\ \cline{1-2} \cline{4-4}
\multirow{2}{*}{2 (29 subjects) } & \multirow{2}{*}{\textcolor{blue}{Placebo}} &  & \multirow{2}{*}{\textcolor{red}{Meloxicam}} &  \\
 &  &  &  &  \\ \hline
\end{tabular}
\end{table}
The objective of the study is to formally test the efficacy of an active drug meloxicam on cats with degenerative joint disease (DJD) in the form of improved daily physical activity counts. Let $\spr{\tau(\cdot,d)}$ be the mean response that is specific to the meloxicam treatment during the $d$th day, since the beginning of the treatment period. We want to formally assess whether, 
\begin{align*}
    H_0: \tau(\cdot,d) \equiv 0  \;\; \forall\; d \;\; \textrm{vs}\;\; H_A: \tau(\cdot,d) \not\equiv  0  \;\; \textrm{for some } d. 
\end{align*}

Most often, especially in a two-period two-treatment crossover design \citep[chapter 2 of][]{jones2014design}, the effect of treatment at period 2 is confounded by the residual effect of the treatments applied in period 1. This residual effect is termed as the ``carryover'' effect of treatment \citep{cochran1941double}. Efficient estimation of direct treatment effect after removing the inherent high between-subject variability, even in the presence of significant carryover, can be done by choosing a suitable crossover design with more than two periods and/or more than two groups, which is the case in our meloxicam study. A significant amount of research has been done in the late twentieth century to address the estimation of direct treatment effect in the presence of carryover \citep{hills1979two}. 

The literature on testing procedures for the bivariate mean function or bivariate smooth effect of a predictor involved in the context of LFD is sparse. Much of the existing methodologies on LFD concentrates on modeling through functional principal component analysis (FPCA) \citep{park2015longitudinal, scheffler2020hybrid}. To the best of our knowledge, the only functional testing procedure developed for LFD tests for invariance of the smooth bivariate mean $\mu(s,d)$ along the longitudinal component $d$ \citep{park2018simple, koner2021profit}. In this paper, we develop a pseudo-generalized F test for the significance of the smooth bivariate effect of treatment on the PA in a hierarchically structured longitudinal functional crossover design. Our methodology relies on projecting $\tau(\cdot,d)$ onto a set of orthogonal basis functions $\spr{\phi_k(\cdot)}_{k \geq 1}$ and testing for the dependence of the coefficient functions over $d$, which we do by extending the generalized F test developed by \cite{wang2012testing} to account for a complex dependence structure. This allows us to transform the global null hypothesis into several simpler hypotheses along the longitudinal direction $d$. Compared to the pseudo-likelihood ratio test (pLRT) developed in \cite{koner2021profit}, the proposed test allows testing for the significance of a smooth bivariate effect, in the presence of other smooth effects in the model. This is crucial in the crossover design, where testing for the direct treatment effect is done in the presence of an overall mean effect and the confounding carryover effect. 

Testing for the direct effect of the treatment in the presence of carryover effect has its challenges. Denoting $\lambda(\cdot,d)$ the carryover effect of the treatment that influences the response in the $d$th day of the washout period. Because the carryover is the residual effect of a treatment, the $H_0$ that $\tau(\cdot,d) = 0$  implies $\lambda(\cdot,d) = 0$. On the other hand, the $H_A$ that $\tau(\cdot,d) \neq 0$ does not require a null carryover effect,  $\lambda(\cdot,d)$. Thus the null and alternative hypotheses of interest can be equivalently written as
\begin{equation} \label{eqn: chap3Nullorig}
    H_0: \tau(s,d) = 0, \lambda(s,d)=0 \;\forall\; s, d \;\;\;\;\text{vs}\;\;\;\; H_A: \tau(s,d) \neq 0 \;\textrm{ for some } s, d.
\end{equation}
Testing the two hypotheses is non-trivial in the sense that the $H_0$ and $H_A$ combined do not span the entire parameter space. Testing a problem of this kind is atypical in the literature as the nuisance parameter $\lambda(s,d)$ is dependent on the actual parameter of interest $\tau(s,d)$. As a result, a generalized F-based statistic considering the null model where both the treatment and the carryover effect are zero, and the full model where both of them are present, will fail to maintain the empirical type I error. To mitigate this problem, we propose a sequential procedure, where at the first stage we test for the significance of the carryover effect in the presence of a treatment effect, and at the second stage, we test for the treatment effect, in a model where the carryover effect is present or absent, depending on the conclusion of the test of carryover conducted at the first stage. The two-stage sequential procedure maintains the type I error, has excellent power in small samples, and is computationally efficient.

The rest of the article is organized as follows. In Section~\ref{sec: chap3method} we formulate the problem and introduce the model framework. The testing procedure along with the theoretical results related to the asymptotic null distribution are described in Section~\ref{sec: chap3prolific} and \ref{sec: chap3prolific_asymp}. Numerical studies are presented in Section~\ref{sec: chap3simstudy} to demonstrate the finite sample performance of the test. Section~\ref{sec: chap3meloxicamstudy} summarizes the findings on the efficacy of the active treatment meloxicam on physical activity, based on the conclusion drawn from the test. Assumptions related to the main theorem of the paper and a detailed proof of the theorems as well as additional results related to the real data applications are provided in the Supplementary Material.

\section{Model Framework} \label{sec: chap3method}

Let the $i$th datum be $\left [G_i,  \left \{d_{ipj}, (Y_{ipjr}, s_{ipjr})_{r=1}^{ R_{ipj}} \right\}_{j, p}, \boldsymbol{C}_{i}\right]$, where $Y_{ipjr}$ is the response for the $i$th subject, measured the $j$th day, within the period $p$ at timepoint of the day $s_{ipjr}$, for all $r=1, \ldots R_{ipj}$, where $j=1, \ldots m_{ip}$ is the day index and $p=1, \ldots, 4$ is the period index. Also $G_i\in\{1,2\}$ is the group to which the $i$th subject is assigned, and $\boldsymbol{C}_{i}$ is an $L$-dimensional vector of covariates measured at baseline. We assume that $\{s_{ipjr}: r=1, \ldots, R_{ipj}\}$ is dense in the compact interval $\mathcal{S}$ for every $i, p, j$, that $m_{ip}$ is small to moderate, 
but the overall set $\spr{d_{ipj} : j=1, \dots, m_{ip}, p =1, \dots, 4, i=1,\dots,n}$ is dense in a compact interval $\mathcal{D}$. In the motivating example $i$ is the cat, $G_i$ is the group at which it has been assigned, $d_{ipj}$ is the actual day of the $j$th observed day relative to the baseline, 
$Y_{ipjr}$ is the PA recorded at the time of the day $s_{ipjr}$. By an abuse of notation we denote by $Y_{ipj}(\cdot)$ the latent trajectory observed for the subject $i$, period $p$, and at the $j$th day such that $Y_{ipj}(s_{ipjr})=Y_{ipjs}$. Without loss of generality, we assume that $R_{ipj} = R$ and $s_{ipjr} = s_r$ for all $r,i,p$ and $j$ and use the index $s$ instead of $s_r$ to denote a typical observation in the entire trajectory.  

In a crossover design, the treatment regime applied at a period $p$ is identified by the group indicator $G_i$; Table~\ref{tab: chap3crossoverlayout} illustrates this for our data application. Let $\mathcal{I}_{ip, \tau}$ be the variable indicating whether the active treatment (meloxicam, for our study) is applied on subject $i$ at period $p$, i.e. $\mathcal{I}_{ip, \tau} = 1$ if $G_i=1, p=1$ or $G_i=2, p=3$ and $\mathcal{I}_{ip, \tau} = 0$ otherwise. Similarly, let $\mathcal{I}_{ip, \lambda}$ be the indicator variable for the carryover effect in the washout period i.e., $\mathcal{I}_{ip, \lambda} = 1$ only if $G_i=1, p=2$ or $G_i=2, p=4$. We model the response at generic time $s$ using a functional additive crossover model (FACM) as,  
\begin{equation} \label{eqn: chap3FACM}
    Y_{ipj}(s) = \mu(s, d_{ipj}) + \tau(s, d_{ipj}) \;\mathcal{I}_{ip, \tau}  + \lambda(s, d_{ipj}) \; \mathcal{I}_{ip, \lambda}  + \sum_{\ell=1}^L C_{i\ell}\beta_\ell(s_r) + \epsilon_{i}(s, d_{ipj}),
\end{equation}
where $\mu(\cdot, d)$ is the population mean, $\tau(\cdot, d)$ is the direct effect of the treatment, and $\lambda(\cdot, d)$ is the carryover effect for the day $d$ since the beginning of the period. Also $C_{i\ell}$'s are the components of the covariate vector $\boldsymbol{C}_i$. Additionally, $\beta_\ell(s)$ quantifies the smooth effect of the baseline covariate $C_{il}$ on the mean response at time $s$. It is assumed that all these population level effects are smooth uni/bivariate functions.  Finally, $\epsilon_{i}(s, d_{ipj})$ is a mean zero random deviation independent and identically distributed across all subjects $i$. The error process is meant to capture the variability of the response trajectory along timpoints, across different days, as well as measurement error. 

\subsection{Alternate formulation of original hypothesis}

We assume FACM framework (\ref{eqn: chap3FACM}).
Let $\spr{\phi_k(s) : s \in \mathcal{S}}_{k \geq 1}$ be a set of orthonormal basis functions in $L^2(\mathcal{S})$, with $\int_\mathcal{S} \phi_k(s) \phi_{k'}(s)ds=\mathrm{I}(k=k')$ for $k, k'\geq 1$. Then, the continuous function  $\tau(s,d)$ can be represented uniquely as $ \tau(s,d) = \sum_{k=1}^{\infty} \tau_k(d)\phi_k(s)$ for all $s \in \mathcal{S} $, where $\tau_k(d) = \int_\mathcal{S} \tau(s,d) \phi_k(s)ds$ is the coefficient function corresponding to $\phi_k(s)$ for $k\geq 1$. Using the same set of orthogonal basis, we can expand the carryover effect as  $ \lambda(s,d) = \sum_{k=1}^{\infty} \lambda_k(d)\phi_k(s)$ with $\lambda_k(d) = \int_\mathcal{S} \lambda(s,d) \phi_k(s)ds$. Then, the original null hypothesis in~(\ref{eqn: chap3Nullorig}) is equivalent to testing a series of simpler hypotheses, 
\begin{equation} \label{eqn: chap3simplerhypink}
    H_{0,k}: \tau_{k}(d) = 0,  \; \lambda_{k}(d) = 0, \; \forall\; d \;\in \mathcal{D} \;\;\;\; \text{vs} \;\;\;\; H_{A,k}: \tau_{k}(d) \neq 0 \text{ for some } d,
\end{equation}
for all $k \geq 1$. 
Thus, we have converted the complex hypothesis testing for the significance of bivariate smooth effect into a series of hypotheses involving univariate functions that are much easier to solve.
Under the null hypothesis $H_{0}$, even though we are interested in testing significance of $\tau_k(d)$, the carryover effect coefficients $\lambda_k(d)$ is also constrained to be zero; these parameters are allowed to vary freely in the alternative hypothesis. In fact, the presence of the carryover effect is only possible if there is a treatment effect. Ideally, we do not want to impose this constraint while estimating $\tau_k(d)$ and $\lambda_k(d)$, as they are separately estimable in our crossover design. Also, assuming a structure of the carryover as a function of the treatment effect would be restrictive \citep[see ][section 1.8]{senn2002cross}. In practice, the carryover effect can be zero while the treatment can be significant. Having a preliminary idea about the presence of carryover can guide us to test the significance of the direct treatment effect in a more substantive model, improving the efficiency of the test. We pursue the testing problem sequentially: we first test for the carryover effect $H_{01, k}: \lambda_k(d) = 0$ versus $H_{A1, k}:  \lambda_k(d)  \neq 0$, and then test for the treatment effect $H_{02, k}: \tau_k(d) = 0$ versus $H_{A2, k}:  \tau_k(d)  \neq 0$ in a model that is informed by the result of the first stage. We present the details in Section~\ref{sec: chap3prolific}. First we set up the framework for testing $H_{01,k}$ and $H_{02,k}$ using the FACM~(\ref{eqn: chap3FACM}) framework.

\subsection{Testing framework under the projected model}\label{sec: chap3testingunderprojectedmodel}

Given the set orthogonal basis functions $\spr{\phi_k(s)}$, consider the projected response,
$Y_{ipj, k} = \int_{\mcS} Y_{ipj}(s)\phi_k(s)ds$, for $k \geq 1$.
The integral can be computed numerically with a very high precision since the grid at which the functional trajectories are observed is dense. The FACM in~(\ref{eqn: chap3FACM}) for the projected response implies the following model for the projected response,
\begin{equation}\label{eqn: chap3kthderivedmodel}
    Y_{ipj, k} = \mu_k(d_{ipj}) + \tau_k(d_{ipj}) \;\mathcal{I}_{ip, \tau}  + \lambda_k(d_{ipj}) \; \mathcal{I}_{ip, \lambda}  + \sum_{\ell=1}^L C_{i\ell}\beta_{\ell,k} + \epsilon_{i,k}(d_{ipj}),
\end{equation}
where the components in the projected model are obtained by projecting each term of the original model onto $\phi_k(\cdot)$, i.e. $\mu_k(d) := \int_{\mcS} \mu(s, d)\phi_k(s)ds$, $\beta_{\ell,k} = \int_{\mcS} \beta_\ell(s)\phi_k(s)ds$, and $\epsilon_{i,k}(d_{ipj}) = \int\epsilon_{i}(s, d_{ipj})\phi_k(s) ds$ is zero-mean residual that is dependent over $p$ and $j$. 

In the context of the meloxicam study, the response trajectories are observed at $20$ days overall, one can approach the hypothesis problem in~(\ref{eqn: chap3simplerhypink}) as testing the nullity of a $20$-dimensional vector $\fpr{\tau_{k}(d_1), \dots, \tau_{k}(d_{20})}^\top $. However, when the time points at which trajectories are observed are very sparse and different for each subject so that the set $\mcD$ can not be construed as a finite set with a small number of elements, or when the variation of $\tau_k(d)$ is smooth over time, standard analysis of variance (ANOVA) type approaches are not powerful.  

For each $k=1, 
\ldots, K$ in part we model the smooth effects in (\ref{eqn: chap3kthderivedmodel}) via truncated polynomial bases and represent them using a linear mixed model (LMM). For example, We expand $\tau_k(d) = \boldsymbol{X} (d)^T \boldsymbol{\alpha}_{\tau,k} + \boldsymbol{Z} (d)^T \boldsymbol{b}_{\tau,k}$, where 
$\boldsymbol{X} (d)$ and $\boldsymbol{Z} (d)$ are the vectors of polynomial basis functions \textcolor{black}{upto order $h_{\tau}$ and splines placed at $Q_{\tau}$ knots respectively, evaluated at $d$, with $\boldsymbol{\alpha}_{\tau,k}$ and $\boldsymbol{b}_{\tau,k}$ being their respective coefficients. The other smooth terms $\mu_k(d)$ (and $\lambda_k(d)$) are similarly represented in terms of polynomial basis of order $h_{\mu}$ (and $h_{\lambda}$) and splines with $Q_{\mu}$ (and $Q_{\lambda}$) knots respectively, with $\boldsymbol{\alpha}_{\tau,k}$ and $\boldsymbol{b}_{\tau,k}$ being their respective coefficients.} Collecting all the terms in a vector format, we can write (\ref{eqn: chap3kthderivedmodel}) as 
\begin{equation}\label{eqn: chap3mixedmodel}
    \M{Y}_{k} = \M{X}_b\V{\alpha}_{b,k} + \M{X}_{\tau}\V{\alpha}_{\tau,k} + \M{X}_{\lambda}\V{\alpha}_{\lambda,k} +
    \M{Z}_{\mu}\V{b}_{\mu, k} + \M{Z}_{\tau}\V{b}_{\tau, k} + \M{Z}_{\lambda}\V{b}_{\lambda, k} + \V{e}_k,
\end{equation}
where $\M{Y}_{k}$ is the vector of $Y_{ipj, k}$'s stacked first over $j$, then over $p$, ad then over $i$'s, $\V{e}_k$ is the vector of residuals, $\V{\alpha}_{\lambda,k}$ and $\V{b}_{\lambda, k}$ are the vectors of the polynomial basis coefficients and the spline coefficients, respectively for $\lambda_{k}(d)$; $\V{\alpha}_{\mu,k}$ and $\V{b}_{\mu, k}$ are defined similarly. It is assumed that $\V{b}_{\mu, k}$, $\V{b}_{\tau, k}$, and $\V{b}_{\lambda, k}$ are independent random variables with zero mean and variances $\sigma^2_{\mu,k}, \sigma^2_{\lambda,k}$, $\sigma^2_{\lambda,k}$, respectively. Also $\V{\alpha}_{\mu\beta,k} = (\V{\alpha}_{\mu,k}^\top, \beta_{1, k}, \dots, \beta_{L, k})^\top$ is a $(h_\mu + L + 1)$ dimensional vector of fixed effect coefficient; the vectors $\V{\alpha}_{\tau,k}$ and $\V{\alpha}_{\lambda,k}$ are modeled as fixed effects too. The details of the above LMM representation are provided in Section~\ref{sec: modelformulation} of the Supplementary Material. In this model framework, $H_{01,k}$ and $H_{02,k}$ can be expressed as
\begin{align*}
    H_{01,k}^\prime :  \V{\alpha}_{\lambda,k} = \V{0}, \; \sigma^2_{\lambda,k} = 0, \;\; \text{vs}\;\; H_{A1,k}^\prime: \V{\alpha}_{\lambda,k} \neq 0 \; \text{or} \; \sigma^2_{\lambda,k} \neq 0, \\
    H_{02,k}^\prime :  \V{\alpha}_{\tau,k} = \V{0}, \; \sigma^2_{\tau,k} = 0, \;\; \text{vs}\;\; H_{A2,k}^\prime: \V{\alpha}_{\tau,k} \neq 0 \; \text{or} \; \sigma^2_{\tau,k} \neq 0.
\end{align*}

Hypothesis testing of smooth effects carried out in a mixed model as above is not new to nonparameteric regression literature; \cite{crainiceanu2004likelihood} developed the null distribution of the restricted likelihood ratio test (RLRT) for testing one variance component. \cite{wang2012testing} developed a generalized F-based testing procedure for testing the significance of a single smooth effect in the presence of multiple variance components. However, Wang and Chen assumed that $\V{e}_k$ are independently distributed across all $i$, $p$, and $j$. In our setup, the covariance structure of $\V{e}_{i,k}$ in~(\ref{eqn: chap3mixedmodel}) is non-trivial. 
We assume an unknown error covariance, $\bSigma_{ik}$ of $\V{e}_{i,k}$. Inspired by \cite{oh2019significance}, we extend the generalized F test to the sequential procedure for testing $H_{0,k}^\prime$ in the form of $H_{01,k}^\prime$ and $H_{02,k}^\prime$, by substituting the true covariance with a proper estimator, as elaborated in our PROjection-based testing for the Lack of Importance of Functional effect in Crossover design (PROLIFIC) in Section~\ref{sec: chap3prolific}.

\section{PROLIFIC: PROjection-based test for Lack of Importance of Function in Crossover design} \label{sec: chap3prolific}

The above testing framework requires a specified set of orthogonal basis system $\spr{\phi_k(\cdot)}_{k\geq 1}$ for the space $\mathcal{L}^2(\mcS)$ to compute the projected response and test $H_{0k}$. We select the eigenfunctions $\spr{\phi_k(\cdot)}_k$ from the spectral decomposition of the so-called ``marginal covariance'' $\Xi(s,\sprime)$ of the error process $\epsilon_i(s, d_{ipj})$, defines by $\Xi(s,\sprime) =\int_\mathcal{D} \epsilon_i(s, t) \epsilon_i(\sprime, t)  g(t) d \, t $, with $g(t)$ accounting for the distribution of the days $t \in\mathcal{D}$; see \cite{park2015longitudinal}. A discussion on the choice of orthogonal bases is provided in Section~\ref{sec: orthogonalbasis} of Supplement. However, the projected response $Y_{ipj, k}$ in model~(\ref{eqn: chap3kthderivedmodel}) is unobserved since the true eigenfunctions $\spr{\phi_k(\cdot)}_{k \geq 1}$ are unknown. Instead, we estimate ${\Xi}(s,\sprime)$ and use the eigenfunctions of the estimated marginal covariance, $\widehat{\phi}_k(\cdot)$, and in turn
 compute the ``quasi projections'' $W_{ipj,k} := \int_{\mcS} Y_{ipj}(s)\widehat{\phi}_k(s)ds$ as a proxy to the unobserved $Y_{ipj.k}$. The high accuracy in the estimation of $\widehat{\phi}_k(\cdot)$ ensures that the quasi projections $W_{ipj,k}$ are sufficiently close to $Y_{ipj,k}$. 

Assuming a completely unknown dependence structure, let  $\sigma^2_k\bSigma_{ik}$ be the $m_{i\centerdot} \times m_{i\centerdot}$ covariance matrix of $\V{e}_{i,k}$ in the linear mixed model~(\ref{eqn: chap3mixedmodel}). Under the independence of the subjects, $\text{Cov}(\V{e}_k) := \sigma^2_k\bSigma_k  = \sigma^2_k\;\text{diag}(\bSigma_{1,k}, \dots, \bSigma_{n,k})$. We use the quasi-projections to construct a consistent estimator of $\bSigma_k$.
We stack $W_{ipj,k}$s for each subject $i$ to construct $\bW_{i,k} := (\bW_{i1,k}^\top,\dots, \bW_{i4,k}^\top)^\top $ with $\bW_{ip,k} := (W_{ip1,k}, \dots, W_{ipm_{ip},k})^\top$. Let $\widehat{\bSigma}_{W,k}$ be a consistent estimator of $\bSigma_{k}$ constructed via $\bW_k := (\bW_{1,k}^\top, \dots, \bW_{n,k}^\top)^\top$. 

We scale the data and design matrices by the inverse square root of $\widehat{\bSigma}_{W,k}$ to compute $\btW_k := \widehat{\bSigma}_{W,k}^{-1/2}\bW_{k}$, $\btX_k := \widehat{\bSigma}_{W,k}^{-1/2}\bX = (\bhSigma_{W,k}^{-1/2}\bX_b, \bhSigma_{W,k}^{-1/2}\bX_\tau, \bhSigma_{W,k}^{-1/2}\bX_\lambda) = (\btX_{b,k}, \btX_{\tau,k}, \btX_{\lambda,k})$ and $\btZ_k := \widehat{\bSigma}_{W,k}^{-1/2}\bZ = (\bhSigma_{W,k}^{-1/2}\bZ_{\mu}, \bhSigma_{W,k}^{-1/2}\bZ_\tau, \bhSigma_{W,k}^{-1/2}\bZ_\lambda) = (\btZ_{\mu,k}, \btZ_{\tau,k}, \btZ_{\lambda,k})$. Moreover, define, $\btX_{-\lambda, k} := (\btX_{b, k}, \btX_{\tau, k})$ to be the subset of $\btX_k$ by removing the columns corresponding to $\balpha_{\lambda,k}$ and similarly, define $\btX_{-\tau, k} := (\btX_{b, k}, \btX_{\lambda, k})$ and the same for $\btZ_{-\lambda,k}$ and $\btZ_{-\tau, k}$. Denote by $r := L+h_\mu+h_\tau + h_\lambda+3$ as the rank of $\bX_k$ and let $\btP_{k} := \btX_k\fpr{\btX_k^\top\btX_k}^{-1}\btX_k^\top$ be the projection matrix onto the column space of $\btX_k$. Throughout the rest of the paper, we assume that $h_{\lambda} < Q_{\lambda} \ll N, h_{\pi} < Q_{\pi} \ll N$, and $h_{\tau} < Q_{\tau} \ll N$, where the symbol ``$\ll$" means that the quantity at its left is considerably smaller than the quantity at its right side. Finally, let $\pi_k := \sigma^2_{\mu, k}/\sigma^2_k $, $\eta_k := \sigma^2_{\tau, k}/\sigma^2_k $ and $\gamma_k := \sigma^2_{\lambda, k}/\sigma^2_k $; the covariance of $\bY_k$ under the mixed model is $\sigma^2_k\bV_k$ with $\bV_k := \bSigma_k + \pi_k \bZ_\mu\bZ_\mu^\top + \eta_k \bZ_\tau\bZ_\tau^\top + \gamma_k \bZ_\lambda\bZ_\lambda^\top$, upto a constant $\sigma^2_k$. Define $\btV_{k}(\pi_k, \eta_k, \gamma_k) := \bI_N + \pi_k\btZ_{\mu,k}\btZ_{\mu,k}^\top + \eta_k\btZ_{\tau,k}\btZ_{\tau,k}^\top + \gamma_k\btZ_{\lambda,k}\btZ_{\lambda,k}^\top$ and $ \btH_{k} = \btH_{k}(\pi_k, \eta_k, \gamma_k) :=  \btX_{k}\fpr{\btX_{k}^\top \btV_{k}(\pi_k, \eta_k, \gamma_k)^{-1}\;\btX_{k}}^{-1} \btX_{k}^\top \btV_{k}(\pi_k, \eta_k, \gamma_k)^{-1} $ to be the generalized projection matrix onto the column space of $\btX_k$ under the full model~(\ref{eqn: chap3mixedmodel}). With this setup, we lay out next the two stages of our testing procedure.

\subsection*{Stage 1: Testing for carryover under the full model}

First, we consider testing $H_{01, k}^{\prime}$ under model~(\ref{eqn: chap3mixedmodel}). The proposed test compares the residual sum of squares (RSS) under the null hypothesis and the alternative hypothesis. The RSS for the full model~(\ref{eqn: chap3mixedmodel}) using the quasi-projections depends on parameters $\pi_k, \eta_k, \gamma_k$, denoted by $qRSS_{k}(\pi_k, \eta_k, \gamma_k)$, can be written as 
$qRSS_{k}(\pi_k, \eta_k, \gamma_k) := \btW_k^\top(\bI_N - \btH_{k})\btV_{k}(\pi_k, \eta_k, \gamma_k)^{-1} (\bI_N - \btH_{k})\btW_k/\sigma^2_k$.
Under the null hypothesis $H_{01,k}^\prime$ we have $\balpha_{\lambda,k} = 0$ and $\gamma_k = 0$; the resulting RSS, denoted by $qRSS_{0,k}^{S1}$ with subscript $0$ for the null hypothesis under stage 1 (superscript $S_1$) becomes
\[
qRSS_{0,k}^{S1}(\pi_k, \eta_k) := \btW_k^\top(\bI_N - \btH_{-\lambda, k})\btV_{k}(\pi_k, \eta_k, 0)^{-1}(\bI_N - \btH_{-\lambda, k})\btW_k/\sigma^2_k
\]
where 
\[
\btH_{-\lambda, k} := \btX_{-\lambda, k}(\btX_{-\lambda, k}^\top\btV_{k}(\pi_k, \eta_k, 0)^{-1} \btX_{-\lambda, k})^{-1}\btX_{-\lambda, k}^\top\btV_{k}(\pi_k, \eta_k, 0)^{-1}.
\]
The pseudo quasi GF (pqGF) statistic for testing $H_{01, k}^{\prime}$ can be constructed as,
\begin{equation}
    pqGF_{N,k}^{S1} := \frac{qRSS_{0,k}^{S1}(\widehat{\pi}_k, \heta_k) - qRSS_{k}(\widehat{\pi}_k, \heta_k, \hgamma_k)}{qRSS_{k}(\widehat{\pi}_k, \heta_k, \hgamma_k)/N}, 
\end{equation}
where $\widehat{\pi}_k$, $\heta_k$, and $\hgamma_k$ are estimated via restricted maximum likelihood (REML) under the full model~(\ref{eqn: chap3mixedmodel}). We call this a pqGF statistic because it is constructed via {\em quasi} projections and the components of the model~(\ref{eqn: chap3mixedmodel}) are scaled by an estimator (hence {\em pseudo}) of the true covariance. The test statistic has a similar form to the test proposed by \cite{wang2012testing}. However, the fundamental difference between the RSS of \cite{wang2012testing} and ${qRSS}_{k}$ is that the latter is constructed through the quasi-data $W_{ipj,k}$, not observed data $Y_{ipj,k}$. This makes the derivation of the null distribution significantly more challenging as $W_{ipj,k}$ are no longer independent across $i$, because $\spr{\widehat\phi_k(\cdot)}_{k \geq 1}$ are obtained from the full data. Nonetheless, the next theorem states that if the eigenfunctions are estimated with high accuracy, then the null distribution changes by a minimal amount. 
\begin{theorem} \label{theorm: chap3carrynull}
Consider the data $\spr{[ d_{ipj}, Y_{ipj}(s), s\in\{s_1, \ldots, s_R\}]_{j=1}^{ m_{ip}}}_{p=1}^4$ for $i=1, \ldots, n$ and suppose that Assumptions~\ref{assump: chap3mibounded}-\ref{assump: chap3epsiloncontinuous} (in Section~\ref{sec: discussionofassump} of the Supplement) hold for the true model (\ref{eqn: chap3FACM}). Let $\lambda^0(s,d)$ be the true carryover effect, and $\lambda^0_k(d) = \int \lambda^0(s,d)\phi_k(s)\,ds$ be the projection of the carryover on the $k$th eigenfunction, that is represented flexibly by a wide class of functions as 
$\lambda^0_k(d) = \sum_{j=0}^{h_\lambda}\alpha^0_{\lambda,k,j}d^j + \sum_{q=1}^{Q_\lambda} b_{\lambda,k,q}(d - \kappa_q)^{h_\lambda}_{+}$ 
where $\{b_{\lambda,k,q}\}_{q=1}^{Q_\lambda} \overset{iid}{\sim} (0, \gamma_k^0\sigma^2_k)$,
for some known value of $\boldsymbol\alpha_{\lambda,k}^0 = (\alpha^0_{\lambda,k,0}, \dots, \alpha^0_{\lambda,k,h_\lambda})^\top$ and $\gamma_k^0 > 0$. Assume that $\sup_{s \in \mathcal{S}}\;\lvert\widehat{\phi}_k(s)-\phi_k(s)\rvert \to 0$ in probability as $n\rightarrow \infty$ and the Assumptions~\ref{assump: chap3normality}-\ref{assump: chap3mineigenvalue} hold for the elements of the projected model in (\ref{eqn: chap3mixedmodel}). Suppose, $\{\xi_{\lambda,k,s}(\pi_k, \eta_k)\}_{s=1}^{Q_\lambda}$ and $\{\varphi_{\lambda,k,s}(\pi_k, \eta_k)\}_{s=1}^{h_\lambda+1}$ are the non-zero eigenvalues of $\btZ_{\lambda,k}^\top\btV_{k}(\pi_k, \eta_k, 0)^{-1}(\bI_N -\btH_{ k}(\pi_k, \eta_k, 0))\btZ_{\lambda,k}$ and $\btZ_{\lambda,k}^\top\btV_{k}(\pi_k, \eta_k, 0)^{-1}(\btH_{ k}(\pi_k, \eta_k, 0) -\btH_{-\lambda, k})\btZ_{\lambda,k}$, respectively. Then for $k=1,\dots,K$, under the null hypothesis 
$H^\prime_{01,k}$, the test statistic
$pqGF_{N,k}^{S1}$ is approximately distributed as the random variable $M_{\lambda,k}(\boldsymbol\alpha_{\lambda,k}^0, \gamma_k^0)$, defined as
\begin{align}
   \frac{\hgamma_k\sum_{s=1}^{Q_\lambda} \frac{\{1 + \gamma_{k}^{0}\xi_{\lambda,k,s}(\widehat{\pi}_k, \heta_k)\}\xi_{\lambda,k,s}(\widehat{\pi}_k, \heta_k)}{1+\hgamma_k\xi_{\lambda,k,s}(\widehat{\pi}_k, \heta_k)}u_s^2 + \sum_{s=1}^{h_\lambda+1} \left(v_s \sqrt{1 + \gamma_k^0 \varphi_{\lambda,k,s}(\widehat{\pi}_k, \heta_k)} + \theta_{\lambda,k,s}(\boldsymbol\alpha_{\lambda,k}^0,\widehat{\pi}_k, \heta_k)\right)^2 }{\frac{1}{N}\spr{\sum_{s=1}^{Q_\lambda} \frac{1 + \gamma_k^{0}\xi_{\lambda,k,s}(\widehat{\pi}_k, \heta_k)}{1+\hgamma_k\xi_{\lambda,k,s}(\widehat{\pi}_k, \heta_k)}u_s^2  + \chi^2_{N-r-Q_\lambda }}}, \label{eqn: chap3asymptnullqpGFcarry}
\end{align}
where 
$\{u_s\}_{s}, \{v_s\}_{s} \overset{iid}{\sim} \textrm{N}(0,1)$, $\chi^2_{N-r-Q_\lambda }$ is a chi-square random variable with degrees of freedom equal to $N-r-Q_\lambda$, and $\theta_{\lambda,k,s}$ is the $s$th element of  $\bG_{\lambda,0,k}^\top\btV_{k}(\pi_k, \eta_k, 0)^{-1/2}\btX_{\lambda,k}\boldsymbol\alpha_{\lambda,k}^0$, where $\bG_{\lambda,0,k}$ is a $N \times (h_\lambda+1)$ matrix with orthonormal columns such that $\bG_{\lambda,0,k}^\top\bG_{\lambda,0,k} = \btV_{k}(\pi_k, \eta_k, 0)^{-1/2}(\btH_{ k}(\pi_k, \eta_k, 0) -\btH_{-\lambda, k})\btV_{k}(\pi_k, \eta_k, 0)^{1/2}$. All random variables in (\ref{eqn: chap3asymptnullqpGFcarry}) are independent. The quantities $\widehat{\pi}_k$, $\heta_k$ and $\hgamma_k$ minimize the spectral decomposition of the negative log-profile restricted likelihood under the full model,
\begin{align*}
    (\widehat{\pi}_k, \heta_k, \hgamma_k)  :=& \argmin_{\pi_k, \eta_k, \gamma_k}\;\; \left[(N-r)\log \spr{\sum_{s=1}^{Q_\lambda} \frac{u_s^2}{1+\gamma_k\xi_{\lambda,k,s}(\pi_k, \eta_k)} + \chi^2_{N-r-Q_\lambda }} \right.  \\
    & \hspace{1 in} +  \left.\sum_{s=1}^{Q_\lambda}\log\spr{1+\gamma_k\xi_{\lambda,k,s}(\pi_k, \eta_k)} + \sum_{s=1}^{Q_\mu + Q_\tau}\log\spr{1+\omega_{-\lambda,k,s}(\pi_k, \eta_k)}\right],
\end{align*}
where $\omega_{-\lambda,k,s}(\pi_k, \eta_k)$ be the $s$th eigenvalue of  $\bD_{-\lambda}(\pi_k, \eta_k)\btZ_{-\lambda,k}^\top(\bI_N -\btP_{ k})\btZ_{-\lambda,k}\bD_{-\lambda}(\pi_k, \eta_k)$ with $\bD_{-\lambda}(\pi_k, \eta_k) := \mathrm{diag}(\sqrt{\pi_k}\;\bI_{Q_\mu}, \sqrt{\eta_k}\;\bI_{Q_\tau}) $.
\end{theorem}

The above theorem presents the distribution of $pqGF_{N,k}^{S1}$ when for any arbitrarily specified value of the true carryover effect. Under the null hypothesis $H_{01,k}^\prime$, the asymptotic null distribution of $pqGF_{N,k}^{S1}$ can be simplified by setting $\boldsymbol\alpha_{\lambda,k}^0$ and $\gamma_k^0$ to zero in the form of $M_{\lambda,k}(\boldsymbol\alpha_{\lambda,k}^0, \gamma_k^0)$ as 
in~(\ref{eqn: chap3asymptnullqpGFcarry}); nonetheless the distribution is non-standard. However one can generate samples efficiently following Algorithm B of \cite{wang2012testing}. The assumption of uniform convergence of $\{\widehat{\phi}_k(s)\}$ is satisfied if the marginal covariance $\Xi(s,\sprime)$ can be estimated at a uniform rate. This can be established following the proof of Theorem 3.1 of \cite{koner2021profit}, provided that the mean of $Y_{ipj}(\cdot)$ is estimated consistently at a uniform rate. Although establishing uniform convergence of the components of FACM~(\ref{eqn: chap3FACM}) is not the focus of this article, there are several works on uniform convergence rate for a nonparametric function such as \cite{delaigle2016nonparametric, xiao2019asymptotics}, which can be directly applied to our case. In that sense, uniform convergence of eigenfunctions is viable in the context of our model~(\ref{eqn: chap3FACM}). A detailed proof is provided in Section~\ref{sec: proofs} of Supplementary Material.

Denote the random variable in~(\ref{eqn: chap3asymptnullqpGFcarry}) after setting $\boldsymbol\alpha_{\lambda, k}^0 = 0$ and $\gamma_k^0 = 0$, as $M_{\lambda,k}(0, 0)$.  Fixing an $\alpha \in (0,1)$, let $M_{\lambda,k,\alpha}(0,0)$ be the $100(1-\alpha)\%$ percentile of the distribution of $M_{\lambda,k}(0, 0)$. The $\alpha$-level test for the null hypothesis $H_{01, k}^{\prime}$ has the rejection region, 
\begin{align}
    \mathcal{R}_{\alpha,k}^{S1} := \spr{\mathcal{Y}_N : pqGF_{N,k}^{S1} \geq M_{\lambda,k,\alpha}(0,0)} \label{eqn: chap3testcarry},
\end{align}
where $\mathcal{Y}_N := \tpr{\spr{Y_{ipj}(s_r): r = 1,\dots,R}: j=1,\dots, m_{ip}, p=1,\dots, 4}_{i=1}^n$ is the collection of response for all subjects and recall $N = \sum_{i=1}^n\sum_{p=1}^4 m_{ip}$.

In the second stage, we test the null hypothesis $H_{02,k}^\prime$ for the direct treatment effect under a model that is evidenced by the conclusion drawn from the first stage. To be specific, if the null hypothesis $H_{01,k}^{\prime}$ is rejected, then we test $H_{02,k}^{\prime}$ under the full model~(\ref{eqn: chap3mixedmodel}) (Stage 2a). However, if we fail to reject $H_{01, k}^{\prime}$ then we test $H_{02,k}^{\prime}$  under a reduced model assuming that there is no carryover effect (Stage 2b). The test statistics along with its null distribution for testing $H_{02,k}^{\prime}$ in the two cases are discussed in turn, next.

\subsection*{Stage 2a: Testing for treatment in the presence of carryover}

Consider testing for a treatment effect $H_{01,k}^\prime$, under the premise that there is a carryover effect. We 
calculate the RSS under $H_{02,k}^\prime$ as $qRSS_{0,k}^{S2a}(\pi_k, \gamma_k) := \btW_k^\top(\bI_N - \btH_{-\tau, k})\btV_{k}(\pi_k, 0, \gamma_k)^{-1}(\bI_N - \btH_{-\tau, k})\btW_k/\sigma^2_k$ with $\btH_{-\tau, k} := \btX_{-\tau, k}(\btX_{-\tau, k}^\top\btV_{k}(\pi_k, 0, \gamma_k)^{-1} \btX_{-\tau, k})^{-1}\btX_{-\tau, k}^\top\btV_{k}(\pi_k, 0, \gamma_k)^{-1}$, since $\balpha_{\tau,k}=0$ and $\eta_k = 0$ under $H_{02,k}^\prime$. Then, the pqGF statistic for testing $H_{02,k}^\prime$ effect under the full model~(\ref{eqn: chap3mixedmodel}) can be constructed as, 
\begin{eqnarray}
pqGF_{N,k}^{S2a} := \frac {
{qRSS}_{0,k}^{S2a}(\widehat{\pi}_k, \hgamma_k) - {qRSS}_{k}(\widehat{\pi}_k, \heta_k, \hgamma_k)}
{{qRSS}_{k}(\widehat{\pi}_k, \heta_k, \hgamma_k)/N},
\nonumber
\end{eqnarray}
where $\widehat{\pi}_k$, $\heta_k$, and $\hgamma_k$ are estimated via REML under the full model. The distribution of $pqGF_{N,k}^{S2a}$ has a similar form as (\ref{eqn: chap3asymptnullqpGFcarry}) 
and is detailed in Eqn.~(\ref{eqn: chap3asymptnullqpGFtrtwcarry}),  Section~\ref{sec: additionalresults} of the Supplement; use notation $M_{\tau,k}(0,0)$ to refer to a random variable with this distribution; the values in parenthesis correspond to setting $\boldsymbol\alpha_{\tau,k}^0 = 0$ and $\eta_k^0 = 0$. As before, fixing an $\alpha \in (0,1)$, let $M_{\tau,k, \alpha}(0,0)$ be the $100(1-\alpha)\%$ percentile of the distribution of $M_{\tau,k}(0,0)$. The $\alpha$-level test for testing the null hypothesis $H_{02, k}^{\prime}$ under the model~(\ref{eqn: chap3mixedmodel}) has rejection region
\begin{equation}
    \mathcal{R}_{\alpha,k}^{S2a} = \spr{\mathcal{Y}_N : pqGF_{N,k}^{S2a} \geq M_{\tau,k, \alpha}(0,0)} \label{eqn: chap3testtrtwcarry}.
\end{equation}
\subsection*{Stage 2b: Testing for treatment in the absence of carryover}

Consider testing for a treatment effect $H_{01,k}^\prime$, under the premise that there is no carryover effect. Testing $H_{02,k}^\prime$ is done under a simpler model that does not a contain carryover effect
\begin{equation}\label{eqn: chap3mixedmodelreduced}
    \M{Y}_{k} = \M{X}_b\V{\alpha}_{b,k} + \M{X}_{\tau}\V{\alpha}_{\tau,k}  +
    \M{Z}_{\mu}\V{b}_{\mu, k} + \M{Z}_{\tau}\V{b}_{\tau, k}   + \V{e}_{k}.
\end{equation}
The RSS under the model~(\ref{eqn: chap3mixedmodelreduced}) is ${qRSS}_{0,k}^{S1}(\pi_k, \eta_k)$, defined in the context of Stage 1. Additionally when $H_{02,k}^\prime$ is true, the RSS simplifies to ${qRSS}_{0,k}^{S2b}(\pi_k) :
= \btW_k^\top(\bI_N - \btH_{b, k})\btV_{k}(\pi_k, 0, 0)^{-1}(\bI_N - \btH_{b, k})\btW_k/\sigma^2_k$ with $\btH_{b, k} := \btX_{b, k}(\btX_{b, k}^\top\btV_{k}(\pi_k, 0, 0)^{-1} \btX_{b, k})^{-1}\btX_{b, k}^\top\btV_{k}(\pi_k, 0, 0)^{-1}$, since $\balpha_{\tau,k} = 0$ and $\eta_k = 0$. Then the pqGF statistic for testing $H_{02,k}^\prime$ under the model~(\ref{eqn: chap3mixedmodelreduced}) can be constructed as $ {pqGF}_{N,k}^{S2b} := N\{{qRSS}_{0, k}^{S2b}(\widehat{\pi}_k) - {qRSS}_{0, k}^{S1}(\widehat{\pi}_k, \heta_k)\}/{qRSS}_{0, k}^{S1}(\widehat{\pi}_k, \heta_k)s $. Fixing an $\alpha \in (0,1)$, let $M_{\tau, -\lambda,k, \alpha}(0,0)$ be the $100(1-\alpha)\%$ percentile of the distribution of the random variable in Eqn.~(\ref{eqn: chap3asymptnullqpGFtrtwocarry}),  Section~\ref{sec: additionalresults} of the Supplement, after setting $\boldsymbol\alpha_{\tau,k}^0 = 0$ and $\eta_k^0 = 0$. A $\alpha$-level test for $H_{02,k}^\prime$ under the model~(\ref{eqn: chap3mixedmodelreduced}) has rejection region
\begin{align}
    \mathcal{R}_{\alpha,k}^{S2b} = \spr{\mathcal{Y}_N : {pqGF}_{N,k}^{S2b} \geq M_{\tau, -\lambda, k,\alpha}(0,0)} \label{eqn: chap3testtrtwocarry}.
\end{align}

\subsection*{\textbf{Combined two-stage test rule for testing $H_{0k}^\prime$}}
We are now ready to present the proposed two-stage test along direction $k$. Fix a level of significance $\alpha_1 \in (0,1)$ for testing the carryover at Stage 1. For any $\alpha \in (0,1)$, a level $\alpha$-test for testing the null hypothesis $H_{0k}^\prime$ has the rejection region,
\begin{equation}\label{eqn: chap3kthtest}
    \mathcal{R}_k(\alpha ; \alpha_1) :=  \spr{\mathcal{R}_{\alpha_1,k}^{S1} \;\bigcap\; \mathcal{R}_{\alpha,k}^{S2a}} \;\bigcup\; \spr{\fpr{{\mathcal{R}^{S1}_{\alpha_1,k}}}^c\;\bigcap\; \mathcal{R}_{\alpha,k}^{S2b}}.
\end{equation}
The test along the direction of $\widehat{\phi}_k(\cdot)$ is the key ingredient of the PROLIFIC, presented in~(\ref{eqn: testruleprolific}).

\section{Asymptotic properties of PROLIFIC} \label{sec: chap3prolific_asymp}

We prove that the test preserves the nominal level and then discuss its power properties. 

\begin{corollary}
Assume the conditions of the Theorem~\ref{theorm: chap3carrynull} and fix a level of significance $\alpha_1 \in (0,1)$ for the test for the carryover effect in~(\ref{eqn: chap3testcarry}). Then for every $k = 1, \dots, K$ and any $\alpha \in (0,1)$, under the null hypothesis $H_0$, $ \mathbb{P}(\mathcal{R}_k(\alpha ; \alpha_1)) \leq \alpha.$
\end{corollary}
The proof follows by the fact that both the tests in~(\ref{eqn: chap3testtrtwcarry}) and ~(\ref{eqn: chap3testtrtwocarry}) are valid. Both the tests at the second stage for the treatment effect must be conducted at the same level, $\alpha$, to ensure that the overall two-stage procedure has an overall size $\alpha$. Moreover, the size of the overall test in~(\ref{eqn: chap3kthtest}) is independent of the level ($\alpha_1$) at which the carryover is tested at Stage 1. This is because under the Gaussian data generating assumption the contrasts contributing to the direct treatment effect and that for the carryover effect are asymptotically independent. Put differently, the inference drawn from the test of carryover effect at Stage 1 does not influence the conclusion from the tests at Stage 2a and 2b, which is also evidenced by the numerical results presented in Section~\ref{sec: chap3simstudy}. This is the main reason behind the widely discussed criticism against the two-stage procedure for the typical AB/BA design \citep[chapter 3]{senn2002cross} as the two tests at Stage 1 and Stage 2 are not independent \citep{freeman1989performance}.

Under the null hypothesis $H_{02,k}^\prime$, both the tests with rejection regions $\mathcal{R}_{\alpha,k}^{S2a}$ and $\mathcal{R}_{\alpha,k}^{S2b}$ are equally capable of making a correct decision about the significance of treatment effect, up to an error level $\alpha$. Thus testing for the carryover effect at the first stage does not have any direct implication when $H_0$ is true. However, the impact of testing for the carryover at the first stage will be profound on the power to detect departure from null. To justify this, suppose that in truth, both the treatment and the carryover effects are significant. Then the power to detect the significant treatment effect using the test in~(\ref{eqn: chap3testtrtwocarry}) will be much smaller than using solely of (\ref{eqn: chap3testtrtwcarry}), as the latter is conducted on a correct model. Therefore, the test for the carryover effect at the first stage provides a tool against possible model misspecification, while testing for treatment effect.

Finally, our projection-based test, PROLIFIC, formally assesses the global null hypothesis $H_0$ by simultaneous testing of $H_{0k}^\prime$ along the $K$ directions and using a Bonferroni multiple testing correction to control for the family-wise error rate. Fix the nominal level $\alpha_1 \in (0,1)$ for the test of carryover at Stage 1. Then, for every $K \geq 1$ and pre-specified nominal level $\alpha$, the rejection region for PROLIFIC is,
\begin{equation}\label{eqn: testruleprolific}
    \mathcal{R}_{\textrm{PROLIFIC}}^K(\alpha;\alpha_1) :=  \bigcup_{k=1}^K\mathcal{R}_k\fpr{\frac{\alpha}{K} ; \frac{\alpha_1}{K}}.
\end{equation}
\begin{corollary}
Assume the setup and the conditions of the Theorem~\ref{theorm: chap3carrynull}. Furthermore, assume that the null hypothesis in~(\ref{eqn: chap3Nullorig}) is true. Then, for any $K \geq 1$ and every $\alpha, \alpha_1 \in (0,1)$: 
\begin{equation*}
    \mathbb{P}\fpr{\mathcal{R}_{\mathrm{PROLIFIC}}^K\fpr{\alpha ; \alpha_1}} \leq \alpha.
\end{equation*}   
\end{corollary}
The choice of the truncation parameter $K$ does not affect the size of the PROLIFIC. However, it affects the power to detect departure from null. In a hypothetical situation, when the treatment effect $\tau(s,d)$ is zero along the eigenfunctions $\spr{\phi_k(\cdot)}_{k=1}^K$, but non-zero along a direction $\phi_j(\cdot)$ for some $j > K$, then the test does not have any power. On the other hand, choosing a large value of $K$ will make the level for the individual hypothesis testing very small, $\alpha/K$ and $\alpha_1/K$, leading to a loss of power. In numerical results, we see that pre-specifying the percentage of variation explained (PVE) to $90\%$, PROLIFIC has the desirable size and strong power performance. 

The choice of nominal level $\alpha_1$ for the test for the carryover at Stage 1 should be determined based on the implication of finding a significant carryover effect. A test for the carryover effect may be important to assess the usefulness of the washout period. A very small value of $\alpha_1$ will lead to poor identification of the carryover, and as a result, we might end up testing for the direct treatment effect in a wrong model. In practice, we recommend choosing a slightly higher value $\alpha_1$ (say $10\%$) compared to $\alpha$ (say $5\%$). The steps associated with implementing PROLIFIC are described in Algorithm~\ref{alg: testingPROLIFIC}. Additional computational details are relegated to Section~\ref{sec: modelestimation} of the Supplement.

\begin{algorithm}[ht]
	\caption{Steps to carry out the test PROLIFIC.} 
	\label{alg: testingPROLIFIC}
Construct smooth estimators of the components of mean model as $\widehat{\mu}_0(s)$, $\widehat{\tau}(s,d)$, $\widehat{\lambda}(s,d)$ and coefficient for other baseline covariates $\spr{\widehat \beta_\ell(s)}_{\ell=1}^L$\;
Compute the demeaned response,
	    $\widetilde{Y}_{ipj}(s_r) := Y_{ipj}(s_r) - \widehat{\mu}_0(s) - \widehat{\tau}(s, d_{ipj}) \;\mathcal{I}_{ip, \tau}  - \widehat{\lambda}(s, d_{ipj}) \; \mathcal{I}_{ip, \lambda}  - \sum_{\ell=1}^L C_{i\ell}\widehat{\beta}_\ell(s)$\;
Obtain a smooth estimator of $\widehat{\Xi}(s,\sprime)$ using the demeaned responses\;
Get $\spr{\widehat{\phi}_k(s)}_{k=1}^K$ from the spectral decomposition of $\widehat{\Xi}(s,\sprime)$ with $K$ chosen by a pre-specified PVE\;
		\For{$k \in \{1, \ldots, K\}$}{
	Construct the projected data $\{[ (d_{ipj}, W_{ipj,k} )_{j=1}^{m_{ip}}, p=1,\dots,4 ], i=1,\dots,n\}$ by calculating $W_{ipj,k} = R^{-1}\sum_{r = 1}^{R} Y_{ipj}(s_r)\widehat{\phi}_k(s_r)$\;
		Compute the p-value $p_k$ of test in~(\ref{eqn: chap3kthtest}) with a specified level of significance  $\alpha_1/K$ for the test of carryover at Stage 1\;
		}
Reject $H_{0}$ in~(\ref{eqn: chap3Nullorig}) if $\min\{p_k : k = 1, \ldots, K\} < \alpha/K$ at some level of significance $\alpha$\;
\end{algorithm}

Next, we consider the power of PROLIFIC. Corollary \ref{corollary: powerPROLIFIC} provides an explicit formula for the asymptotic power of PROLIFIC when at least one of the projections of $\tau(s,d)$ onto the first $K$ eigenfunctions
is non-zero. When $\tau(s, d) \neq 0$, and $K$ is sufficiently large, it is unlikely that all of its projections on the first $K$ directions are equal to zero. 

\begin{corollary} \label{corollary: powerPROLIFIC}
Consider the data $\spr{[ d_{ipj}, Y_{ipj}(s), s\in\{s_1, \ldots, s_R\}]_{j=1}^{ m_{ip}}}_{p=1}^4$ for $i=1, \ldots, n$ and that Assumptions~\ref{assump: chap3mibounded}-\ref{assump: chap3epsiloncontinuous} hold for the true model (\ref{eqn: chap3FACM}). Assume that the eigenfunctions are consistently estimated, i.e. $\sup_{s \in \mathcal{S}}\;\lvert\widehat{\phi}_k(s)-\phi_k(s)\rvert \to 0$ in probability as $n\rightarrow \infty$ and the Assumptions~\ref{assump: chap3normality}-\ref{assump: chap3mineigenvalue} hold for the elements of the projected model in (\ref{eqn: chap3mixedmodel}). Suppose that we want to test the null hypothesis under the alternative $H_1: \tau(s,d) = \tau^0(s,d)$ where $\tau_0(s,d)$ is a smooth bivariate function of $s$ and $d$ with $\lVert \tau^0\rVert > 0$, and that the true carryover effect is $\lambda^0(s,d)$,  which can be either zero or non-zero. Then, for any fixed truncation $K$, and specified significance level $\alpha \in (0,1)$, the power function of PROLIFIC is

{\small\begin{align*}
\mathbb{P}\,\left( \mathcal{R}^K_{\textrm{PROLIFIC}} \mid H_{1} \right) 
\geq &\max_{k =1,\dots,K}  \left\{\mathbb{P}\left(M_{\lambda, k}(\boldsymbol\alpha_{\lambda,k}^0,\gamma_k^0) \geq M_{\lambda,  k,\alpha_1/K}(0,0)\right) \mathbb{P}\left(M_{\tau, k}(\boldsymbol\alpha_{\tau,k}^0,\eta_k^0) \geq M_{\tau,  k,\alpha/K}(0,0)\right) \right. \\
&+ \left.\mathbb{P}\left(M_{\lambda, k}(\boldsymbol\alpha_{\lambda,k}^0,\gamma_k^0) < M_{\lambda,  k,\alpha_1/K}(0,0)\right)\mathbb{P}\left(M_{\tau, -\lambda, k}(\boldsymbol\alpha_{\tau,k}^0,\eta_k^0) \geq M_{\tau, -\lambda, k,\alpha/K}(0,0)\right)\right\},
\end{align*}}
where 
$\alpha_1 \in (0,1)$ is the significance level used for testing the carryover effect at stage 1, and the quantities $\{(\boldsymbol\alpha_{\tau,k}^0,\eta_k^0)\}_{k=1}^K$ are the coefficients involved in the truncated polynomial basis representation of the projection of $\tau^0(s,d)$ on the first $K$-eigenbasis of the covariance function, as mathematically represented by 
$\tau^0_k(d) = \int \tau^0(s, d)\phi_k(s) \,ds$, with 
\begin{equation*} 
\tau^0_k(d) = \sum_{j=0}^{h_\tau}\alpha^0_{\tau,k,j}d^j + \sum_{q=1}^{Q_\tau} b_{\tau,k,q}(d - \kappa_q)^{h_\tau}_{+}, \qquad  \{b_{\tau,k,q}\}_{q=1}^{Q_\tau} \overset{iid}{\sim} (0, \eta_k^0\sigma^2_k),
\end{equation*}
for some non-zero fixed $\boldsymbol\alpha_{\tau,k}^0 = (\alpha^0_{\tau,k,0}, \dots, \alpha^0_{\tau,k,h_\tau})^\top$ and $\eta_k^0 > 0$. Similarly $\boldsymbol\alpha_{\lambda,k}^0$ and/or $\gamma_k^0$ are quantities involved in the truncated polynomial basis representation 
of the projection of the true carryover effect on the $k$th eigenfunction, $\lambda^0_k(d) = \int \lambda^0(s, d)\phi_k(s) \,ds$ as in Theorem \ref{theorm: chap3carrynull}.
\end{corollary}

\textcolor{black}{Corollary~\ref{corollary: powerPROLIFIC} presents the power of the test to reject the null hypothesis for a fixed sample size $n$. It is important to note that, the distribution of $M_{\lambda, k}(\boldsymbol\alpha_{\lambda,k}^0,\gamma_k^0)$, $M_{\tau, k}(\boldsymbol\alpha_{\tau,k}^0,\eta_k^0)$, and $M_{\tau, -\lambda, k}(\boldsymbol\alpha_{\tau,k}^0,\eta_k^0)$ goes to $\infty$ as the sample size $n \to \infty$. This ensures that for a fixed non-zero true treatment effect the power of PROLIFIC goes to $1$ as the sample size goes to infinity.} 

\section{Simulation study}\label{sec: chap3simstudy}
\textbf{Data generation}:
To assess the performance of PROLIFIC, we generate data for sample size $n$ varying from $100$ to $300$. As described, we consider a crossover design with $4$ periods and within each period the response profiles are observed sparsely over $m_{ip}$ time points. The number of profiles per subject in each period, $m_{ip}$, is generated randomly from $\{8, 9, \ldots, 12\}$ (\textit{low} sparsity level). For each $m_{ip}$, the time points $d_{ipj}$ are uniformly sampled from $\mathcal{D} =[0,1]$. Each profile is observed over a grid of $R=101$ equispaced points in $\mathcal{S} =[0,1]$. The model is 
\begin{align*}
     Y_{ipj}(s) = \mu(s, d_{ipj}) + \tau(s, d_{ipj}) \;\mathcal{I}_{ip, \tau}  + \lambda(s, d_{ipj}) \; \mathcal{I}_{ip, \lambda}  + \epsilon_{i}(s, d_{ipj}).
\end{align*}

The structure of the mean model are: $\mu(s, d) = 2d\cos(\pi s/2)$, $\tau(s, d) = \delta \cos(\pi s/2)(1 + 4f_\beta(0.8d, a, b))$ and $\lambda(s, d) = \delta \gamma \cos(\pi s/2)(1 + 4f_\beta(2d/3 + 0.8, a, b))$ with $a < b$ and $f_\beta(x, a, b)$ is the density of Beta distribution with parameter $a > 0$ and $b > 0$. The structure for the treatment effect ensures that the projection of $\tau(s,d)$ on $\phi_k(s)$, $\int_0^1 \tau(s,d)\phi_k(s) ds$ is proportional to the $f_\beta(x, a, b)$ which is right-skewed for $a < b$, $k=1,2$. From a practical perspective (as we also see it in the data analysis), this \textit{going up trend and vanishing feature} nature of treatment effect is reasonable. We scale the time point by $0.8$ to ensure that $\tau(s,d)$ is not zero at the end of the period represented by $d=1$. Furthermore, the assumed structure of the carryover can be viewed as a continuation of the treatment effect in the next period which is non-zero for $d < 0.3$ and then it vanishes for $d \geq 0.3$. The knob $\delta$ controls the magnitude of the treatment and the carryover. Both $\tau(s,d)$ and $\lambda(s,d)$ are equal to zero if $\delta = 0$. On the other hand, $\gamma$ controls the magnitude of carryover relative to the treatment. Setting $\delta > 0$ and $\gamma = 0$, we can enforce the absence of carryover even when the treatment is significant. We take the parameters of the beta density as $a = 2, b = 4$. The error process $\epsilon_i(s,d)$ has a mean zero and complex covariance described in Section~\ref{sec: additionalsimresults} of  Supplement.

\textbf{Remarks on competing methods}:
After the data are projected along directions $\phi_k(\cdot)$, other hypothesis testing approaches can be used to test the significance of a smooth function. For example, if the null hypothesis $H_{02,k}: \tau_k(d) = 0$ is of interest , then one alternative is the $L_2$-norm based statistic $T_{ZC,k} = \int_{0}^1 \spr{\widehat{\tau}_k(t) - \tau_k(t)}^2 dt $ constructed by \cite{zhang2007statistical} to test for null hypothesis $H_{0k}$. When the functional data are observed densely, the asymptotic null distribution of the test statistic is a mixture of chi-square distributions with weights corresponding to the eigenvalue of the covariance matrix $\bSigma_k$ and can be approximated either by replacing the unknown values of the mixture components or by Bootstrap. When applying this approach to our situation, the functions are obtained as projections on data-driven directions; we call this method an adapted version of the test and abbreviate it as Ad-ZC. The conclusion about the overall test is obtained by applying the same two-stage procedure implemented for PROLIFIC.

\subsection*{Assessing performance of the PROLIFIC test}

\begin{table}[pos=ht]
\centering
\caption{Empirical size of \textbf{\em PROLIFIC} based on 5000 simulations}
\label{tab: size_prolific}
\resizebox{\textwidth}{!}{
\begin{tabular}{lclcccccc}
\hline
\hline
 & & & &$\alpha = 0.01$&$\alpha = 0.05$&$\alpha = 0.1$ & $\alpha = 0.15$\\
\hline
$n = 50$ & & $\alpha_1 = 0.05$ & & 0.012 (0.002) & 0.054 (0.003) & 0.106 (0.004) & 0.154 (0.005)\\
   & & $\alpha_1 = 0.1$ & & 0.013 (0.002) & 0.055 (0.003) & 0.106 (0.004) & 0.155 (0.005)\\ 

  $n = 100$ & & $\alpha_1 = 0.05$ & & 0.012 (0.002) & 0.049 (0.003) & 0.097 (0.004) & 0.144 (0.005)\\  
   & &$\alpha_1 = 0.1$ & & 0.012 (0.002) & 0.049 (0.003) & 0.097 (0.004) & 0.144 (0.005)\\  
\hline
\hline
\end{tabular}}
\end{table}

\begin{figure}[pos=ht]
    \centering
        \subfloat[True carryover is non-zero]{\includegraphics[scale = 0.21]{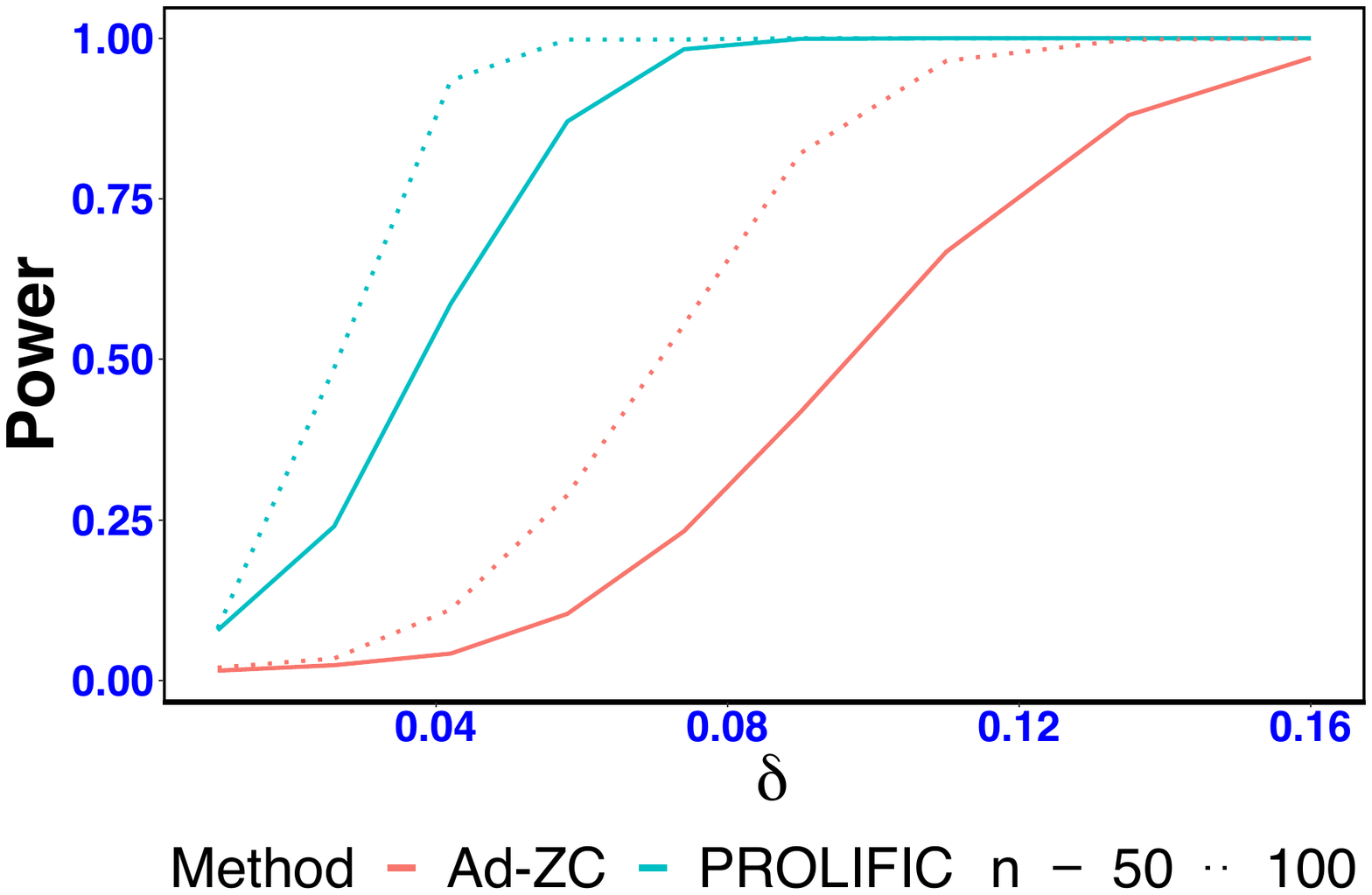}}\qquad
        \subfloat[True carryover is zero]{\includegraphics[scale=0.21]{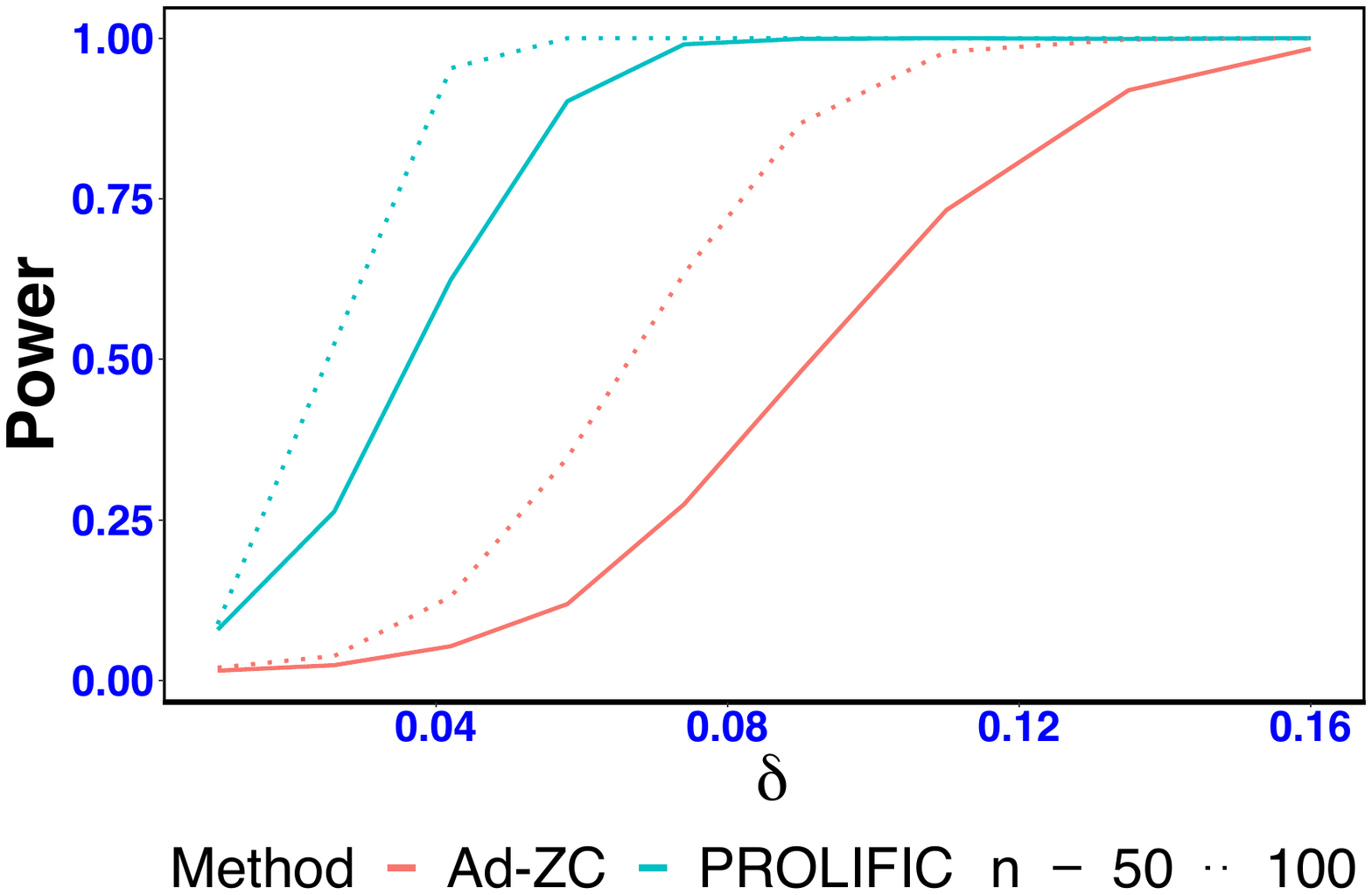}}
        \newline
    \subfloat[True carryover is non-zero]{\includegraphics[scale = 0.21]{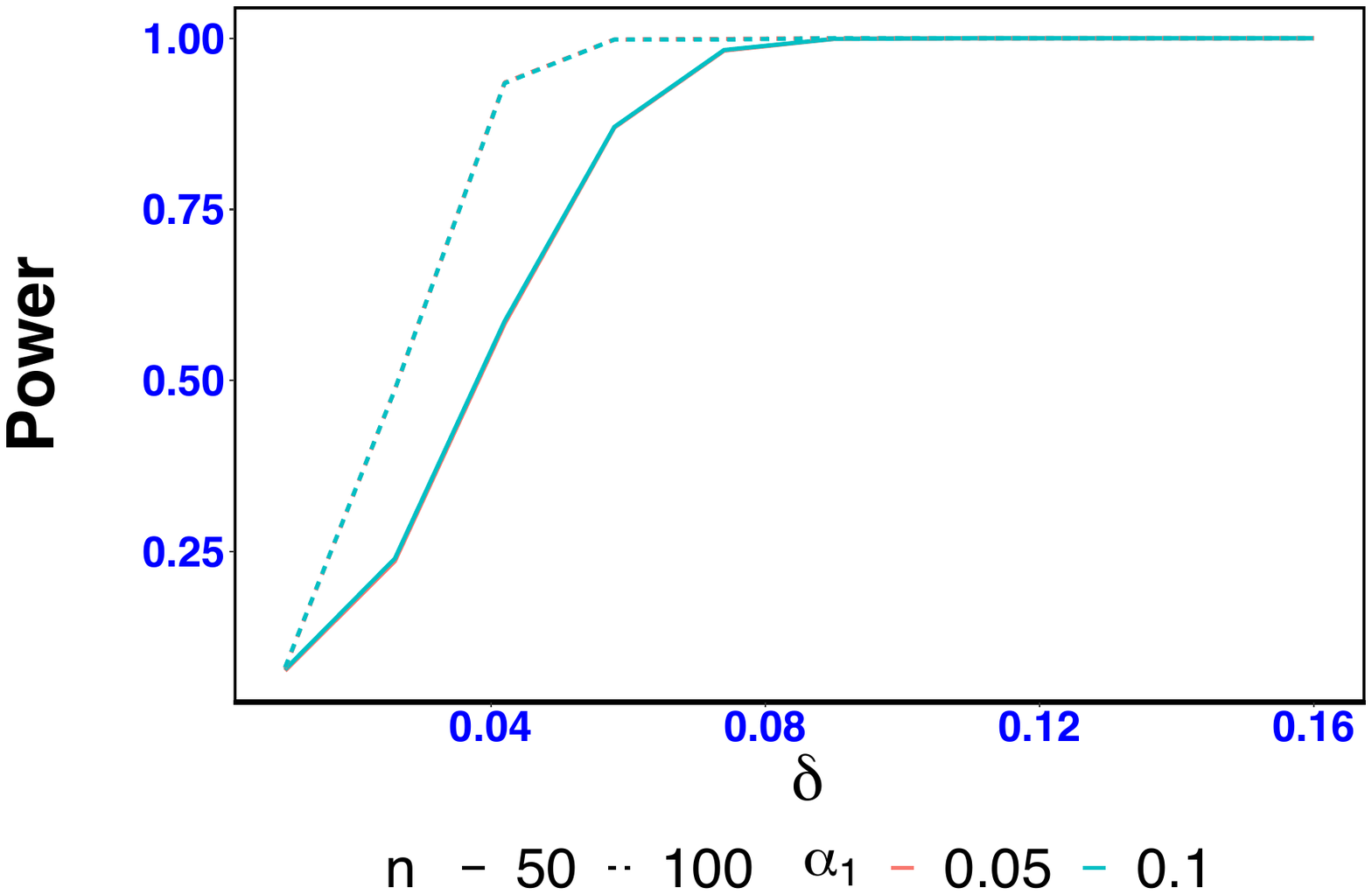}}\qquad
        \subfloat[True carryover is zero]{\includegraphics[scale=0.21]{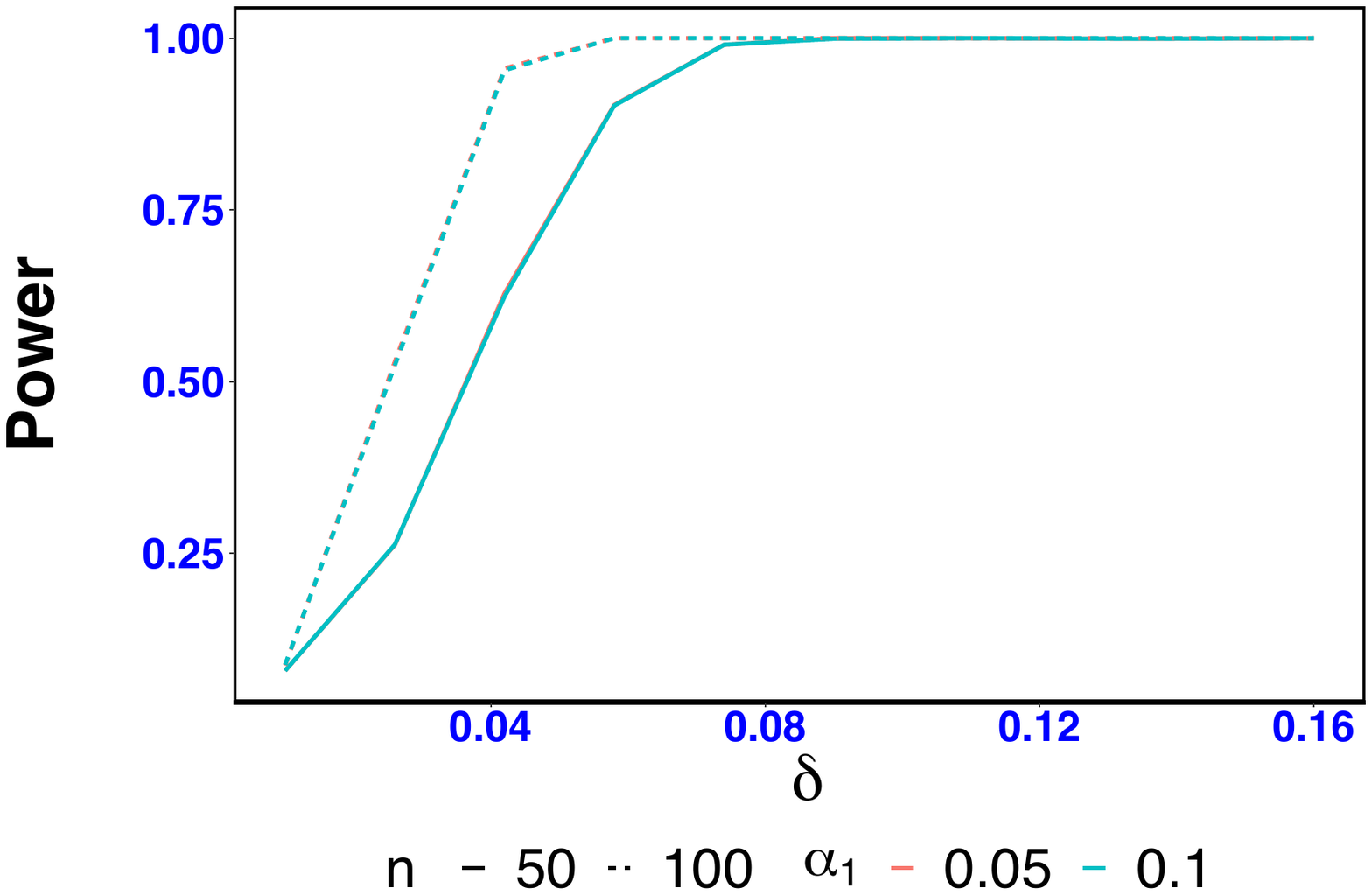}}
    \caption{\textit{Upper panel}: Power curve of the PROLIFIC as a function of $\delta$ for $\alpha = 0.05$ in comparison to the power of Ad-ZC (when null distribution is approximated by bootstrap) across $n = 50$ and $100$, when the true $\lambda(s,d)$ is (a) nonzero and (b) zero, based on $1000$ simulations. The level of the carryover test $\alpha_1$ is set at $0.1$. \textit{Lower panel}: Power curve of PROLIFIC for $\alpha = 0.05$ across sample size $n = 50$ and $100$ and two different levels of the test of carryover at Stage $1$, $\alpha_1 = 0.05$ and $0.1$, when the true $\lambda(s,d)$ is (c) nonzero and (d) zero.}
    \label{fig: prolificpowercurve}
\end{figure}
\textbf{Size:} The empirical type I error rate of the PROLIFIC across small ($n=50$) and medium ($n = 100$) sample size in 5000 simulations is presented in Table~\ref{tab: size_prolific} at specified nominal levels $\alpha = 0.01, 0.05, 0.1$ and $0.15$, with the two different levels of the carryover test at the first stage ($\alpha_1$). 
Even for sample sizes $n = 50$, the empirical size of PROLIFIC is maintained within twice the standard error of the stipulated nominal level. The results confirm that the size of the overall test is not influenced by the level of significance for the carryover effect ($\alpha_1$). Ad-ZC test with the null distribution approximated by a mixture of chi-squares fails to maintain the nominal size even for sample sizes of $n=100$. When its null distribution is approximated with Bootstrap, the performance is conservative; see Table \ref{tab: size_zc} for details. 

\textbf{Power:} Figure \ref{fig: prolificpowercurve} depicts the empirical power of PROLIFIC as a function of $\delta$, that quantifies the departure from the null, for small and medium sample size, when the level of significance is fixed at $\alpha = 0.05$. Results are based on $1000$ simulations. The left panels show the power when the carryover effect is non-zero, while the right panels plot the power when there is no carryover effect. The upper panel demonstrates that PROLIFIC is more powerful than Ad-ZC to detect departure from null, irrespective of whether carryover is zero or not. 
The bottom panels show that the power of  PROLIFIC is not affected by the significance level of the test for carryover $\alpha_1$; the overlapped plots correspond to $\alpha_1=0.05$ and $\alpha_1=0.10$. 

The strength of PROLIFIC in detecting a slight departure from null even with a small sample size can be attributed to the fact that both the treatment and carryover contrasts are estimable after removing the subject variation. 
Overall, the numerical results testify to the effectiveness of the two-stage procedure in detecting the significant direct treatment effect in a crossover design, when both the treatment and carryover contrasts are separately estimable, in contrast to the widely criticized lack of power of the two-stage procedure in the case of AB/BA crossover design \citep[][chapter 3]{senn2002cross}.

\section{Meloxicam study of cats with osteoarthritis} \label{sec: chap3meloxicamstudy}

The motivating data application is a meloxicam study of $58$ household cats with osteoarthritis. The cats were enrolled in a completely randomized double masked placebo-controlled crossover trial conducted at the College of Veterinary Medicine of North Carolina State University. The subjects were randomized into two groups. As described in Table~\ref{tab: chap3crossoverlayout}, the first group received a single dose of the active drug meloxicam for 20 days in the first period, followed by a placebo during the last three periods. Whereas, the second group received meloxicam at period $3$ and received placebo at all the remaining three periods. For each cat, additional information is collected at baseline, such as age, weight, and a radiologist assessed degenerative joint disease (DJD) score that evaluates the disease severity. The objective of the study is to understand the efficacy of an active drug meloxicam on joint pain as reflected by a change in the PA counts, measured at every minute level during the day by an activity monitor. See \cite{gruen2015criterion} for complete details of the study. 

\textbf{Data preprocessing}:
Figure~\ref{fig: rawandrollact}(a) presents the daily raw physical activity (PA) counts recorded by accelerometer for a randomly selected cat over $5$ days in every period. To stabilize the high variability in the activity profiles induced by sudden changes (such as jumps, large movement, etc.) in the cats' activity, we log-transform the raw PA counts, followed by taking a cumulative average at every minute in the day, and denote the response variable as 
$
Y_{ipj}(s_r) := r^{-1}\sum_{\ell = 1}^{r} \log(1+\text{PA}_{ipj}(s_\ell))
$, where $r=1, \dots, 1440$ denotes the minutes in a day with $r = 1$ referring to the midnight (12:00:00 AM) and $\text{PA}_{ipj}(s_\ell)$ is the PA counts for the $i$th at the $\ell$th minute of the $j$th day in the $p$th period. We focus on the time of the day $5$AM to $10$PM, when the owners are more likely to be awake. Figure \ref{fig: logcumactid1-2} in Supplement shows the cumulative average of the logPA for four selected cats over different days in the four periods. 
We model the logPA during 5 AM to 10 PM for cat $i$, observed the $j$th time within period $p$ 
as $Y_{ipj}(\cdot)$.

The number of PA profiles in each period ($m_{ip}$) varies across cats and a frequency distribution of $\spr{m_{ip}}_{i=1}^n$ for all four periods is provided in Figure~\ref{fig: freqmipandboxplots}(a) in the Supplement. The age of cats in the study varies between 6 years to 21 years with a median age of 12 years. Two cats, who are either very old or very young compared to the remaining cohort are removed from the analysis; see Figure~\ref{fig: freqmipandboxplots}(b)) in the Supplementary Material. 

\subsection{Data analysis}
To test for the significance of the treatment effect, we posit the FACM,
{\begin{align*} 
    \nonumber Y_{ipj}(s) &= \mu(s, \text{Age}_{ipj}) + \tau(s, d_{ipj}) \;\mathcal{I}_{ip, \tau}  + \lambda(s, d_{ipj}) \; \mathcal{I}_{ip, \lambda}  \\
     & \hspace{1.5 in} + \text{WT}_{i}\beta_1(s) + \text{WE}_{ipj}\beta_2(s) + \text{DJD}_i\beta_3(s) + \epsilon_{i}(s, d_{ipj}), 
\end{align*}
}
where $ \text{Age}_{ipj}$ is the age (days) of the $i$-th cat at the $j$-th day of the $p$-th period, $WT$ is the cat's weight (grams)
$\text{WE}_{ipj}$ is the weekend indicator, i.e. it takes values $1$ if the $j$-th day in the $p$-th period is a weekend, otherwise it is zero. All the other components in the model are defined previously. Here $\text{Age}_{ipj}$ was computed by adding the cat's baseline age to the number of days elapsed in the study period. The DJD score is a factor that is expected to affect the PA. As indoor cats' activity can be influenced by their owner's presence at home, it is important to distinguish between a workday and a weekend day.

\begin{figure}[pos=tbp]
    \centering
        \subfloat[Univariate cross-section of $\widehat{\tau}(\cdot,d)$]{\includegraphics[scale = 0.4]{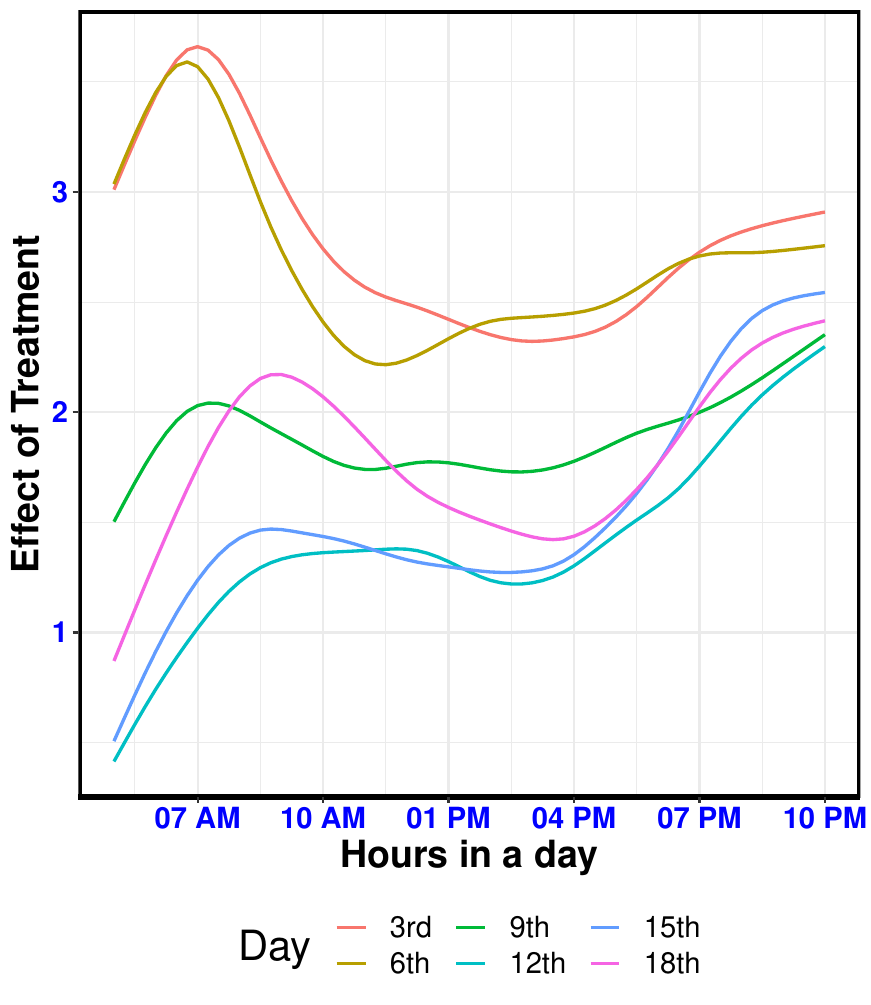}}\qquad
        \subfloat[Univariate cross-section of $\widehat{\lambda}(\cdot,d)$]{\includegraphics[scale=0.4]{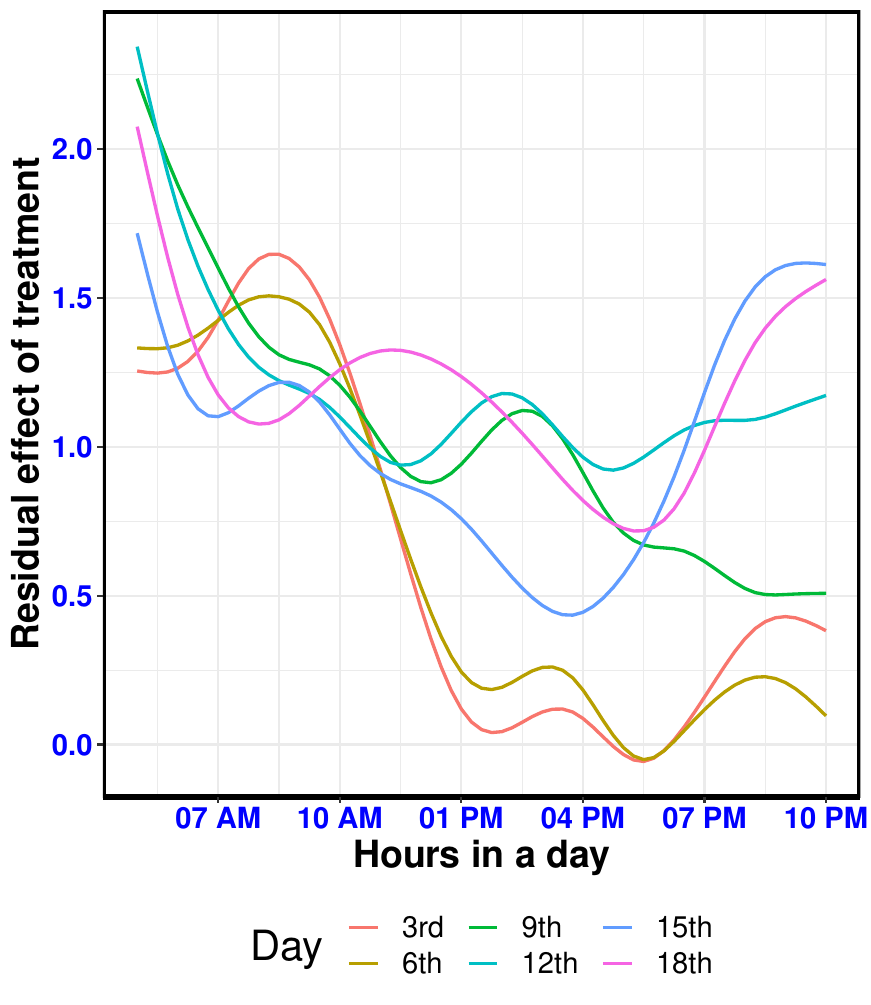}}
        \newline
    \subfloat[Estimated correlation function]{\includegraphics[scale = 0.4]{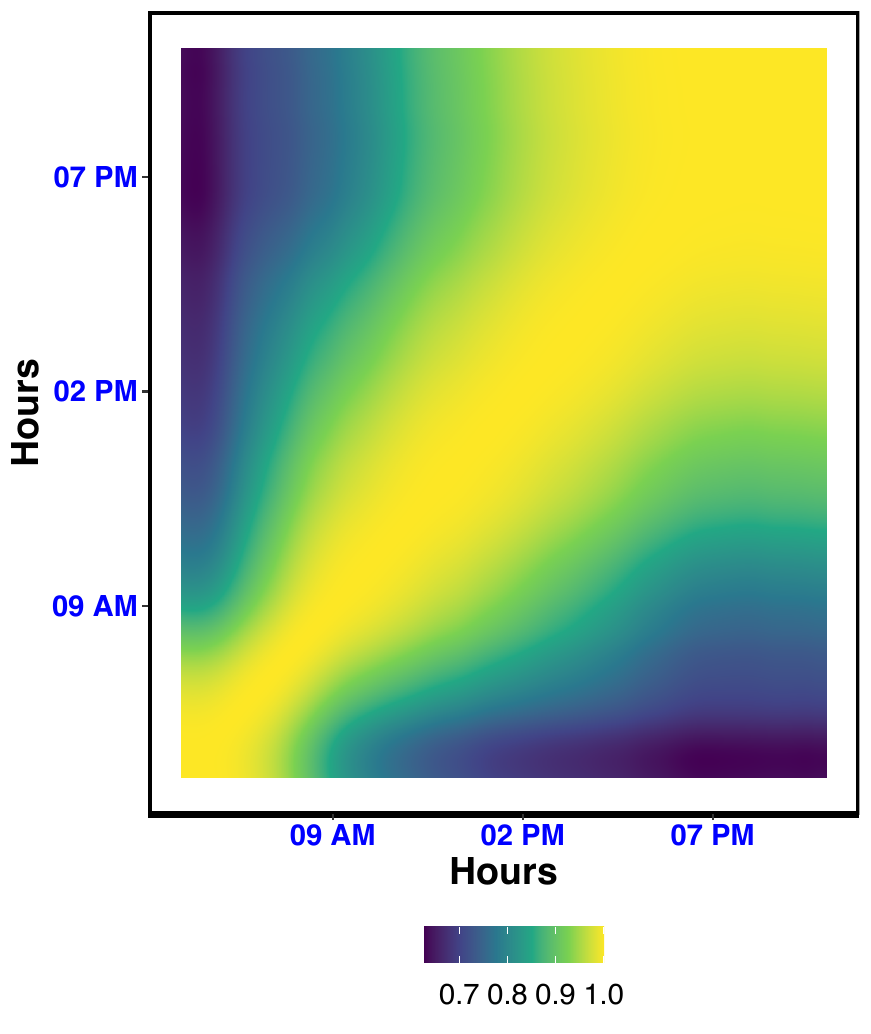}}\qquad
        \subfloat[Estimated eigenfunctions, $\widehat{\phi}_k(s)$]{\includegraphics[scale=0.4]{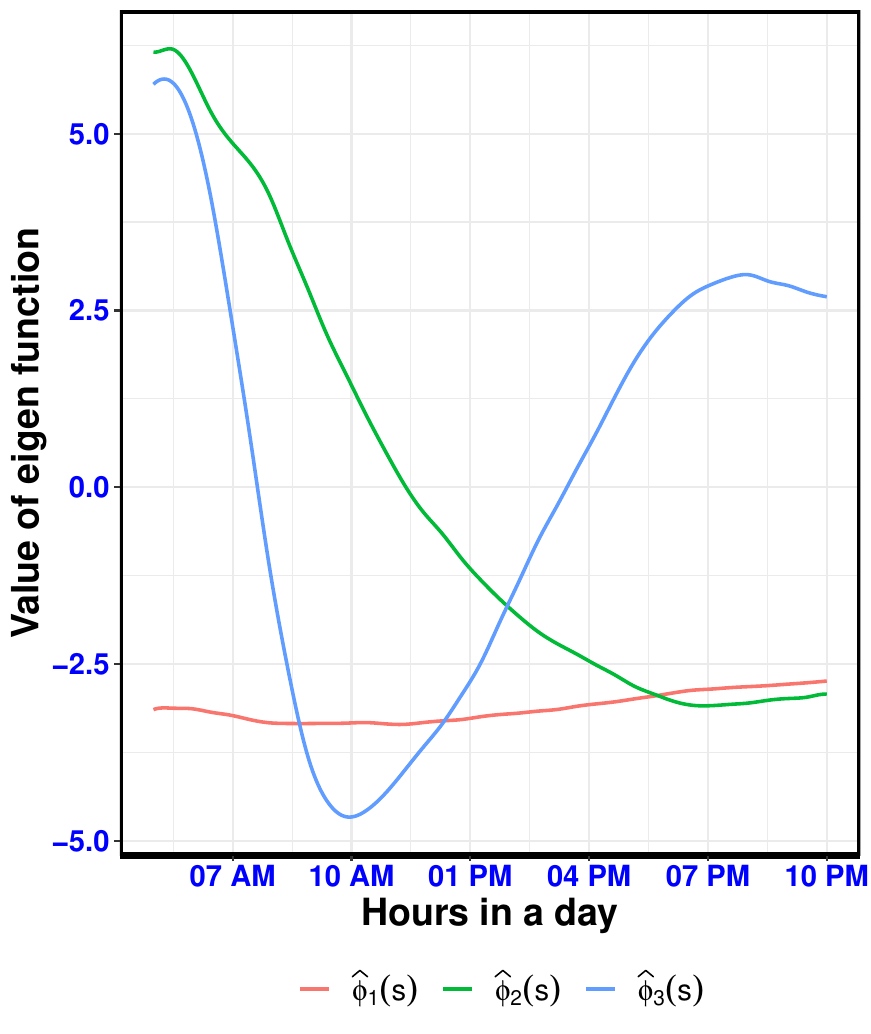}}
    \caption{\textit{Upper panel}: (a) Estimated treatment $\widehat{\tau}(\cdot,d)$ and (b) carryover $\widehat{\lambda}(\cdot,d)$ for $3$rd, $6$th, $9$th, $12$th, $15$th, and $18$th day in the period. The numbers in the y-axis are multiplied by $100$. \textit{Lower panel}: (c) Estimated marginal correlation function obtained from $\widehat{\Xi}(s,\sprime)$, as a bivariate function of the hours in the day. (d) Estimated eigenfunctions $\widehat{\phi}_k(s)$, $k=1,2,3$ as a function of hours in the day, obtained by specifying a PVE $= 95\%$.}
    \label{fig:Estcomponent}
\end{figure}

To address the question of interest we test for the significance of the direct treatment effect $H_0: \tau(s,d) = 0$ vs $H_1: \tau(s,d) \neq 0$ for some $s$ and $d$, using PROLIFIC as described in Section~\ref{sec: chap3prolific}. We estimate the bivariate smooth functions in the model such as $\mu(s, \text{Age}_{ipj})$, and $\tau(s, d_{ipj})$, nonparametrically using a tensor product of cubic spline basis via \verb|gam()| function in the \verb|mgcv| package \citep{wood2004stable} in R. Figure~\ref{fig:Estcomponent}, top panels show the estimates of the treatment and the carryover effects over $6$ equidistant days in a period; they indicate a higher effect of treatment in the first half of the period. Details about implementation and inference about baseline covariates are provided in Section~\ref{sec: additionaldataresults} of the Supplementary Material.

After estimating the fixed components of the model, we demean the response and estimate the marginal covariance function $\widehat{\Xi}(s,\sprime)$ via sandwich smoother. The spectral decomposition yields $K = 3$ eigenfunctions $\{\widehat{\phi}_k(\cdot)\}$ explaining $95\%$ of the total variation. The estimated marginal correlation along with the eigenfunctions are presented in the lower panel of Figure~\ref{fig:Estcomponent}. The growing correlation along the center is a direct consequence of the cumulative average of activities, described in the preprocessing step. Using the $\{\widehat{\phi}_k(\cdot)\}$, we obtain the projected response $W_{ipj,k} =  \sum_{r=1}^R Y_{ipj}(s_r)\widehat{\phi}_k(s_r) $, whose model is described by $W_{ipj,k} = \mu_k(\text{Age}_{ipj}) + \tau_k(d_{ipj}) \;\mathcal{I}_{ip, \tau}  + \lambda_k(d_{ipj}) \; \mathcal{I}_{ip, \lambda}  + \text{WT}_{i}\beta_{1,k} + \text{WE}_{ipj}\beta_{2,k}+ \text{DJD}_i\beta_{3,k}$.
The framework of PROLIFIC allows us to test for the $H_{0,k}: \tau_k(d) = 0, \lambda_k(d) = 0$ vs $H_{1,k}: \tau_k(d) \neq 0$, $k=1,2,3$ simultaneously under the projected model. We model the smooth components in the model using a truncated linear basis and apply the two-stage testing procedure. For each of $k = 1,2,3$, we test for the significance of a carryover effect and obtain the p-values $0.99, 0.17$, and $0.41$
suggesting no evidence of the presence of residual effect of the treatment during the washout period. Next, we test for the significance of the treatment effect following the test rule in Stage 2b, dropping the carryover term from the projected model. The p-values of the three significance tests for the treatment are $< 0.0001$, $0.06$, and $0.11$, suggesting strong evidence for the significance of the direct effect of meloxicam. The results are coherent with the conclusion based on the p-values $(0.01, 0.55, 0.41)$ of the Ad-ZC test when the null distribution is approximated by $5000$ bootstrap samples. The relatively higher p-values reflect the conservative nature of the Ad-ZC test to detect departure from null, compared to the more powerful PROLIFIC, that we observed in Section~\ref{sec: chap3simstudy}.

\section*{Software}
\label{sec: software}

Software in the form of R code for the entire simulation study is available publicly on 
\url{https://github.com/SalilKoner/ProjectionTesting}. 
\section*{Supplementary Material}
\label{sec: supp}
The reader is referred to the on-line Supplementary Material for detailed derivation of the null and alternate distribution of the test statistics and additional data analysis results.

\section*{Funding}
The authors would like to acknowledge the support of NSF, grant number DMS 1454942.

\bibliographystyle{cas-model2-names}

\bibliography{refs_main}

\begin{thebibliography}{24}
\expandafter\ifx\csname natexlab\endcsname\relax\def\natexlab#1{#1}\fi
\providecommand{\url}[1]{\texttt{#1}}
\providecommand{\href}[2]{#2}
\providecommand{\path}[1]{#1}
\providecommand{\DOIprefix}{doi:}
\providecommand{\ArXivprefix}{arXiv:}
\providecommand{\URLprefix}{URL: }
\providecommand{\Pubmedprefix}{pmid:}
\providecommand{\doi}[1]{\href{http://dx.doi.org/#1}{\path{#1}}}
\providecommand{\Pubmed}[1]{\href{pmid:#1}{\path{#1}}}
\providecommand{\bibinfo}[2]{#2}
\ifx\xfnm\relax \def\xfnm[#1]{\unskip,\space#1}\fi
\bibitem[{Bussmann et~al.(2001)Bussmann, Martens, Tulen, Schasfoort, Van Den
  Berg-Emons and Stam}]{bussmann2001measuring}
\bibinfo{author}{Bussmann, J.}, \bibinfo{author}{Martens, W.},
  \bibinfo{author}{Tulen, J.}, \bibinfo{author}{Schasfoort, F.},
  \bibinfo{author}{Van Den Berg-Emons, H.}, \bibinfo{author}{Stam, H.},
  \bibinfo{year}{2001}.
\newblock \bibinfo{title}{Measuring daily behavior using ambulatory
  accelerometry: the activity monitor}.
\newblock \bibinfo{journal}{Behavior Research Methods, Instruments, \&
  Computers} \bibinfo{volume}{33}, \bibinfo{pages}{349--356}.
\bibitem[{Cochran et~al.(1941)Cochran, Autrey and Cannon}]{cochran1941double}
\bibinfo{author}{Cochran, W.}, \bibinfo{author}{Autrey, K.},
  \bibinfo{author}{Cannon, C.}, \bibinfo{year}{1941}.
\newblock \bibinfo{title}{A double change-over design for dairy cattle feeding
  experiments}.
\newblock \bibinfo{journal}{Journal of Dairy Science} \bibinfo{volume}{24},
  \bibinfo{pages}{937--951}.
\bibitem[{Crainiceanu and Ruppert(2004)}]{crainiceanu2004likelihood}
\bibinfo{author}{Crainiceanu, C.M.}, \bibinfo{author}{Ruppert, D.},
  \bibinfo{year}{2004}.
\newblock \bibinfo{title}{Likelihood ratio tests in linear mixed models with
  one variance component}.
\newblock \bibinfo{journal}{Journal of the Royal Statistical Society: Series B}
  \bibinfo{volume}{66}, \bibinfo{pages}{165--185}.
\bibitem[{Delaigle et~al.(2016)Delaigle, Hall and
  Zhou}]{delaigle2016nonparametric}
\bibinfo{author}{Delaigle, A.}, \bibinfo{author}{Hall, P.},
  \bibinfo{author}{Zhou, W.X.}, \bibinfo{year}{2016}.
\newblock \bibinfo{title}{Nonparametric covariate-adjusted regression}.
\newblock \bibinfo{journal}{The Annals of Statistics} \bibinfo{volume}{44},
  \bibinfo{pages}{2190--2220}.
\bibitem[{Di et~al.(2009)Di, Crainiceanu, Caffo and Punjabi}]{di2009multilevel}
\bibinfo{author}{Di, C.Z.}, \bibinfo{author}{Crainiceanu, C.M.},
  \bibinfo{author}{Caffo, B.S.}, \bibinfo{author}{Punjabi, N.M.},
  \bibinfo{year}{2009}.
\newblock \bibinfo{title}{Multilevel functional principal component analysis}.
\newblock \bibinfo{journal}{The Annals of Applied Statistics}
  \bibinfo{volume}{3}, \bibinfo{pages}{458}.
\bibitem[{Freeman(1989)}]{freeman1989performance}
\bibinfo{author}{Freeman, P.}, \bibinfo{year}{1989}.
\newblock \bibinfo{title}{The performance of the two-stage analysis of
  two-treatment, two-period crossover trials}.
\newblock \bibinfo{journal}{Statistics in medicine} \bibinfo{volume}{8},
  \bibinfo{pages}{1421--1432}.
\bibitem[{Goldsmith et~al.(2015)Goldsmith, Zipunnikov and
  Schrack}]{goldsmith2015generalized}
\bibinfo{author}{Goldsmith, J.}, \bibinfo{author}{Zipunnikov, V.},
  \bibinfo{author}{Schrack, J.}, \bibinfo{year}{2015}.
\newblock \bibinfo{title}{Generalized multilevel function-on-scalar regression
  and principal component analysis}.
\newblock \bibinfo{journal}{Biometrics} \bibinfo{volume}{71},
  \bibinfo{pages}{344--353}.
\bibitem[{Gruen et~al.(2015)Gruen, Griffith, Thomson, Simpson and
  Lascelles}]{gruen2015criterion}
\bibinfo{author}{Gruen, M.E.}, \bibinfo{author}{Griffith, E.H.},
  \bibinfo{author}{Thomson, A.E.}, \bibinfo{author}{Simpson, W.},
  \bibinfo{author}{Lascelles, B.D.X.}, \bibinfo{year}{2015}.
\newblock \bibinfo{title}{Criterion validation testing of clinical metrology
  instruments for measuring degenerative joint disease associated mobility
  impairment in cats}.
\newblock \bibinfo{journal}{PLoS One} \bibinfo{volume}{10},
  \bibinfo{pages}{e0131839}.
\bibitem[{Hills and Armitage(1979)}]{hills1979two}
\bibinfo{author}{Hills, M.}, \bibinfo{author}{Armitage, P.},
  \bibinfo{year}{1979}.
\newblock \bibinfo{title}{The two-period cross-over clinical trial.}
\newblock \bibinfo{journal}{British journal of clinical pharmacology}
  \bibinfo{volume}{8}, \bibinfo{pages}{7--20}.
\bibitem[{Jones and Kenward(2014)}]{jones2014design}
\bibinfo{author}{Jones, B.}, \bibinfo{author}{Kenward, M.G.},
  \bibinfo{year}{2014}.
\newblock \bibinfo{title}{Design and analysis of cross-over trials}.
\newblock \bibinfo{publisher}{Chapman and Hall/CRC}.
\bibitem[{Koner et~al.(2024)Koner, Park and Staicu}]{koner2021profit}
\bibinfo{author}{Koner, S.}, \bibinfo{author}{Park, S.Y.},
  \bibinfo{author}{Staicu, A.M.}, \bibinfo{year}{2024}.
\newblock \bibinfo{title}{Profit: projection-based test in longitudinal
  functional data}.
\newblock \bibinfo{journal}{Journal of Nonparametric Statistics} ,
  \bibinfo{pages}{1–28}\URLprefix
  \url{http://dx.doi.org/10.1080/10485252.2023.2294885},
  \DOIprefix\doi{10.1080/10485252.2023.2294885}.
\bibitem[{Oh(2019)}]{oh2019significance}
\bibinfo{author}{Oh, S.}, \bibinfo{year}{2019}.
\newblock \bibinfo{title}{Significance tests for longitudinal functional data.}
\newblock \URLprefix
  \url{https://repository.lib.ncsu.edu/items/14d6d3d7-dceb-42a0-9ec4-dc1e70484c31}.
  \bibinfo{note}{ph.D. Thesis}.
\bibitem[{Park and Staicu(2015)}]{park2015longitudinal}
\bibinfo{author}{Park, S.Y.}, \bibinfo{author}{Staicu, A.M.},
  \bibinfo{year}{2015}.
\newblock \bibinfo{title}{Longitudinal functional data analysis}.
\newblock \bibinfo{journal}{Stat} \bibinfo{volume}{4},
  \bibinfo{pages}{212--226}.
\bibitem[{Park et~al.(2018)Park, Staicu, Xiao and Crainiceanu}]{park2018simple}
\bibinfo{author}{Park, S.Y.}, \bibinfo{author}{Staicu, A.M.},
  \bibinfo{author}{Xiao, L.}, \bibinfo{author}{Crainiceanu, C.M.},
  \bibinfo{year}{2018}.
\newblock \bibinfo{title}{Simple fixed-effects inference for complex functional
  models}.
\newblock \bibinfo{journal}{Biostatistics} \bibinfo{volume}{19},
  \bibinfo{pages}{137--152}.
\bibitem[{Ratkowsky et~al.(1992)Ratkowsky, Alldredge and
  Evans}]{ratkowsky1992cross}
\bibinfo{author}{Ratkowsky, D.}, \bibinfo{author}{Alldredge, R.},
  \bibinfo{author}{Evans, M.A.}, \bibinfo{year}{1992}.
\newblock \bibinfo{title}{Cross-over experiments: design, analysis and
  application}. volume \bibinfo{volume}{135}.
\newblock \bibinfo{publisher}{CRC Press}.
\bibitem[{Reider et~al.(2020)Reider, Bai, Scharfstein, Zipunnikov,
  Investigators et~al.}]{reider2020methods}
\bibinfo{author}{Reider, L.}, \bibinfo{author}{Bai, J.},
  \bibinfo{author}{Scharfstein, D.O.}, \bibinfo{author}{Zipunnikov, V.},
  \bibinfo{author}{Investigators, M.O.S.}, et~al., \bibinfo{year}{2020}.
\newblock \bibinfo{title}{Methods for step count data: Determining “valid”
  days and quantifying fragmentation of walking bouts}.
\newblock \bibinfo{journal}{Gait \& Posture} \bibinfo{volume}{81},
  \bibinfo{pages}{205--212}.
\bibitem[{Scheffler et~al.(2020)Scheffler, Telesca, Li, Sugar, Distefano, Jeste
  and {\c{S}}ent{\"u}rk}]{scheffler2020hybrid}
\bibinfo{author}{Scheffler, A.}, \bibinfo{author}{Telesca, D.},
  \bibinfo{author}{Li, Q.}, \bibinfo{author}{Sugar, C.A.},
  \bibinfo{author}{Distefano, C.}, \bibinfo{author}{Jeste, S.},
  \bibinfo{author}{{\c{S}}ent{\"u}rk, D.}, \bibinfo{year}{2020}.
\newblock \bibinfo{title}{Hybrid principal components analysis for
  region-referenced longitudinal functional eeg data}.
\newblock \bibinfo{journal}{Biostatistics} \bibinfo{volume}{21},
  \bibinfo{pages}{139--157}.
\bibitem[{Senn(2002)}]{senn2002cross}
\bibinfo{author}{Senn, S.}, \bibinfo{year}{2002}.
\newblock \bibinfo{title}{Cross-over trials in clinical research}.
  volume~\bibinfo{volume}{5}.
\newblock \bibinfo{publisher}{John Wiley \& Sons}.
\bibitem[{Wang and Chen(2012)}]{wang2012testing}
\bibinfo{author}{Wang, Y.}, \bibinfo{author}{Chen, H.}, \bibinfo{year}{2012}.
\newblock \bibinfo{title}{On testing an unspecified function through a linear
  mixed effects model with multiple variance components}.
\newblock \bibinfo{journal}{Biometrics} \bibinfo{volume}{68},
  \bibinfo{pages}{1113--1125}.
\bibitem[{Wood(2004)}]{wood2004stable}
\bibinfo{author}{Wood, S.N.}, \bibinfo{year}{2004}.
\newblock \bibinfo{title}{Stable and efficient multiple smoothing parameter
  estimation for generalized additive models}.
\newblock \bibinfo{journal}{Journal of the American Statistical Association}
  \bibinfo{volume}{99}, \bibinfo{pages}{673--686}.
\bibitem[{Xiao(2019)}]{xiao2019asymptotics}
\bibinfo{author}{Xiao, L.}, \bibinfo{year}{2019}.
\newblock \bibinfo{title}{Asymptotics of bivariate penalised splines}.
\newblock \bibinfo{journal}{Journal of Nonparametric Statistics}
  \bibinfo{volume}{31}, \bibinfo{pages}{289--314}.
\bibitem[{Xiao et~al.(2015)Xiao, Huang, Schrack, Ferrucci, Zipunnikov and
  Crainiceanu}]{xiao2015quantifying}
\bibinfo{author}{Xiao, L.}, \bibinfo{author}{Huang, L.},
  \bibinfo{author}{Schrack, J.A.}, \bibinfo{author}{Ferrucci, L.},
  \bibinfo{author}{Zipunnikov, V.}, \bibinfo{author}{Crainiceanu, C.M.},
  \bibinfo{year}{2015}.
\newblock \bibinfo{title}{Quantifying the lifetime circadian rhythm of physical
  activity: a covariate-dependent functional approach}.
\newblock \bibinfo{journal}{Biostatistics} \bibinfo{volume}{16},
  \bibinfo{pages}{352--367}.
\bibitem[{Zhang and Chen(2007)}]{zhang2007statistical}
\bibinfo{author}{Zhang, J.T.}, \bibinfo{author}{Chen, J.},
  \bibinfo{year}{2007}.
\newblock \bibinfo{title}{Statistical inferences for functional data}.
\newblock \bibinfo{journal}{The Annals of Statistics} \bibinfo{volume}{35},
  \bibinfo{pages}{1052--1079}.
\bibitem[{Zhang et~al.(2019)Zhang, Li, Keadle, Matthews and
  Carroll}]{zhang2019review}
\bibinfo{author}{Zhang, Y.}, \bibinfo{author}{Li, H.}, \bibinfo{author}{Keadle,
  S.K.}, \bibinfo{author}{Matthews, C.E.}, \bibinfo{author}{Carroll, R.J.},
  \bibinfo{year}{2019}.
\newblock \bibinfo{title}{A review of statistical analyses on physical activity
  data collected from accelerometers}.
\newblock \bibinfo{journal}{Statistics in biosciences} \bibinfo{volume}{11},
  \bibinfo{pages}{465--476}.

\end{thebibliography}

\end{document}